\documentclass[final, aps, pra, floatfix, reprint, nofootinbib, citeautoscript, superscriptaddress, balancelastpage, longbibliography, showkeys]{revtex4-2}
\usepackage[T1]{fontenc}
\usepackage[english]{babel}

\usepackage{amsmath}
\usepackage{amssymb}
\usepackage{bm}
\usepackage{wasysym}
\usepackage{textcomp}

\usepackage[normalem]{ulem}
\usepackage{tabularx}
\usepackage[colorlinks, urlcolor=blue, citecolor=blue, linkcolor=blue, pdfstartview=FitH]{hyperref}
\usepackage[all]{hypcap}
\usepackage[dvipsnames]{xcolor}
\usepackage{subfigure}
\usepackage{graphicx}

\def\affiIOFFE{Ioffe\ Institute, 194021 St.~Petersburg, Russia}

\begin{document}

\author{M.\ V.\ Petrenko}
\affiliation{\affiIOFFE}

\author{A.\ S.\ Pazgalev}
\affiliation{\affiIOFFE}

\author{A.\ K.\ Vershovskii}
\email{antver@mail.ioffe.ru}
\affiliation{\affiIOFFE}

\selectlanguage{English}

\title{Linear dichroism signals due to the alignment in the ground state of hydrogen-like atoms and their inversion in a single-beam optical pumping scheme}
\begin{abstract}
We present the results of theoretical and experimental investigation of optical linear dichroism signals caused by alignment, and arising under the action of pumping with linearly polarized light in a magnetic field transverse to the pumping in an ensemble of hydrogen-like (including alkali) atoms. Several simple models implying mixing of Zeeman sublevels of the excited state are considered, and common patterns among them are identified. It is shown that in the general case the properties of the medium causing dichroism can be characterized by only two parameters of the longitudinal and transverse linear polarizability, while in the case of complete mixing – by a single linear polarizability parameter. Expressions that relate these parameters to the transition probabilities are provided, as is an extension of the theory for the case of incomplete mixing of sublevels of the excited state. The single-beam scheme for recording linear dichroism signals, which is of greatest interest from a practical point of view, is investigated. It is shown that incomplete mixing can lead to inversion of dichroism signals in systems including more than one optical transition. A comparison with the results of an experimental study of dichroism in alkali metal (cesium) vapors is made. \end{abstract}

\keywords{linear dichroism, optical alignment, linearly polarized light, transverse magnetic field, Voigt geometry, balanced signal detection}

\maketitle

\section{Introduction}\label{sec:1}

The study of spin effects arising in atomic media, as well as in their closest analogues, such as color centers in crystals, quantum dots, etc., under the influence of optical pumping is currently acquiring increasing importance. Optical pumping is widely used in quantum magnetometry, including biological applications 
\cite{Budker_Romalis_2007,
Romalis_2022, 
Boto_Holmes_Leggett_Roberts_Shah_Meyer_Munoz_Mullinger_Tierney_Bestmann_2018, Knappe_Schwindt_Shah_Hollberg_Kitching_Liew_Moreland_2005,
Alem_Hughes_Buard_Cheung_Maydew_Griesshammer_Holloway_Park_Lechuga_Coolidge_2023}, 
atomic clocks 
\cite{Knappe_Schwindt_Shah_Hollberg_Kitching_Liew_Moreland_2005,
Ludlow_Boyd_Ye_Peik_Schmidt_2015}, 
nuclear magnetic gyroscopes 
\cite{Meyer_Larsen_2014,
Sorensen_Thrasher_Walker_2020}, 
quantum interferometers 
\cite{Tino_2021,
Tino_Kasevich_2014}, 
quantum information 
\cite{Paraiso_Woodward_Marangon_Lovic_Yuan_Shields_2021},
methods of laser cooling of atoms 
\cite{Schreck_Druten_2021},
and many other fields. Two-photon methods such as coherent population trapping, electro-induced transparency, etc. are being successfully developed, and can also be classified as optical pumping \cite{Nagel_Graf_Naumov_Mariotti_Biancalana_Meschede_Wynands_1998,
Vanier_2005,
Fleischhauer_Imamoglu_Marangos_2005}. 
The vast majority of these studies are devoted to the effects of optical orientation -- a process in which atoms, under the influence of circularly polarized radiation, acquire a non-zero collective magnetic moment. 

This article is devoted to the specific features of optical alignment. In the quantum-mechanical representation, alignment 
\cite{Fourcault_Romain_Gal_Bertrand_Josselin_Prado_Labyt_Palacios-Laloy_2021,
Meraki_Elson_Ho_Akbar_Kozbial_Kolodynski_Jensen_2023,
Wang_Wu_Xiao_Wang_Peng_Guo_2021,
Breschi_Weis_2012,
Rochester_Ledbetter_Zigdon_Wilson-Gordon_Budker_2012,
LeGal_Lieb_Beato_Jager_Gilles_Palacios-Laloy_2019,
LeGal_Palacios-Laloy_2022,
Budker_Gawlik_Kimball_Rochester_Yashchuk_Weis_2002} 
corresponds to a symmetric distribution of atoms over the projections of the magnetic moment, or magnetic (Zeeman) sublevels, -- in contrast to orientation, which corresponds to an asymmetric distribution of atoms over magnetic sublevels. Alignment effect (a more rigorous definition of which can be found in 
\cite{Blum_2012,
Omont_1977}), which is not accompanied by the emergence of a magnetic moment in the medium, has not yet found such wide application. 

However, it has a very significant advantage over the orientation effects that occur when pumping with circularly polarized light: alignment is most effectively induced by linearly polarized light, with the alignment parameters being determined by the angle between the plane of light polarization and the magnetic field vector. 

Accordingly, in the standard Voigt geometry, where the pump beam is perpendicular to the magnetic field vector, it becomes possible to vary the angle between the light polarization plane and the direction of the field, which is impossible when pumping with circularly polarized light. This, in turn, makes it possible to control the alignment in the system and to record the alignment-induced linear dichroism (LD) signals that are not related to magnetic resonance. This possibility was experimentally investigated in 
\cite{Petrenko_Pazgalev_Vershovskii_2024}, where a method for stabilizing the laser light frequency without modulating it was proposed based on the LD effects. Instead of the laser frequency, the direction of the magnetic field vector is modulated in the scheme proposed in 
\cite{Petrenko_Pazgalev_Vershovskii_2024}, which makes it possible to preserve the spectral characteristics of the laser light.

Theoretical analysis and experimental study of the alignment of LD signals under the influence of resonant pumping with linearly polarized light in alkali metal (cesium) vapors are presented in 
\cite{Meraki_Elson_Ho_Akbar_Kozbial_Kolodynski_Jensen_2023}. 
In 
\cite{Fomin_Kozlov_Petrov_Smirnov_Petrenko_Zapasskii_2025}, a study of the angular dependences of LD signals arising in a single-beam scheme in a medium in which anisotropy arises under the influence of optical pumping is presented. It was also shown in 
\cite{Fomin_Kozlov_Petrov_Smirnov_Petrenko_Zapasskii_2025}
that even in the absence of nonlinearity, such signals can be characterized by the “dipole'' or “quadrupole'' nature of the angular dependence. These two main types of angular dependences differ in the number of local maxima of the signal (two and four, respectively), when the relative angle between the field and the plane of polarization of light changes from 0 to 2$\pi $.

\begin{figure*}[!t]  
	\includegraphics[width=\linewidth]{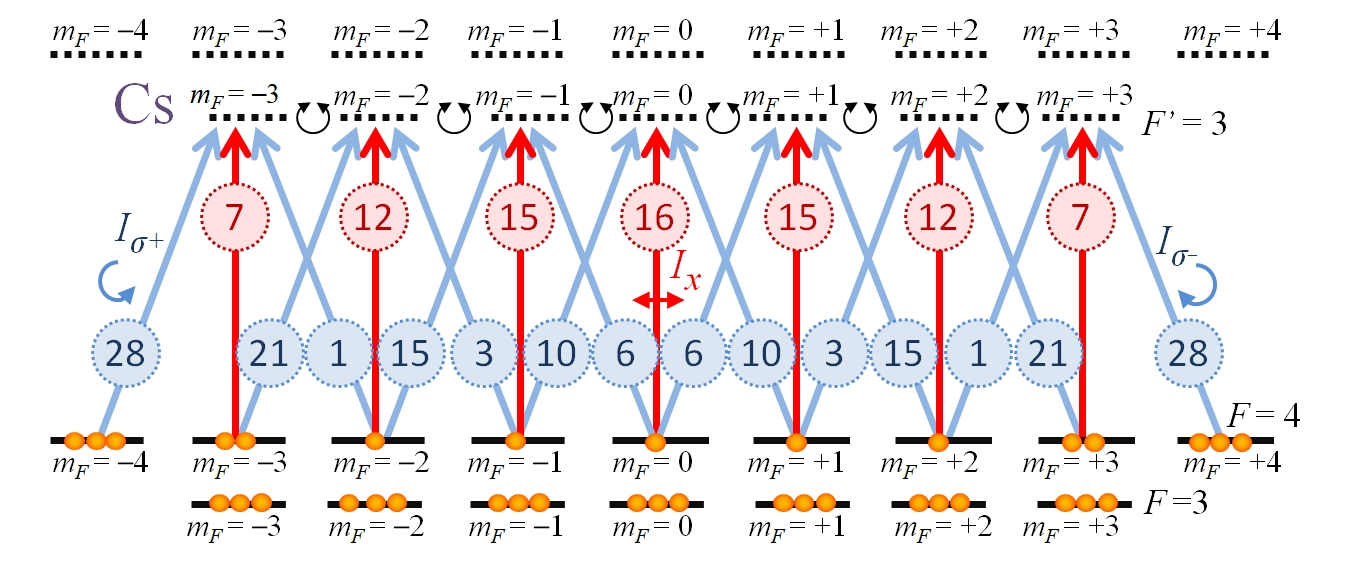}
	\caption{Model of Cs pumped by linearly polarized light of the D\textsubscript{1} line, resonant to the transition $F = 4 \leftrightarrow F' = 3$. Notations: $I_x = I_0 \cos^2(\varphi)$, $Iy = I_0 \sin^2(\varphi)$ are the projections of the intensity of the linearly polarized pump light onto the axes specified by the direction of the magnetic field ($\mathbf{B}||\mathbf{x}$); intensities $I_{\sigma +}$, $I_{\sigma -}$ ($I_{\sigma +} = I_{\sigma -} = I_y/2$) are the result of decomposition of the linearly polarized component of the pump light $I_y$ into circularly polarized components.}\label{fig1}
\end{figure*}

Our work was inspired by 
\cite{Fomin_Kozlov_Petrov_Smirnov_Petrenko_Zapasskii_2025}. In our work, we specify the concepts of “dipole'' and “quadrupole'' dichroism in an ensemble of single-electron (hydrogen-like) atoms under the influence of the linearly polarized light in the Voigt geometry (the field is directed perpendicularly to the pump beam). We show that under weak pumping LD in such media is always described by “dipole'' dependencies, and they become “quadrupole'' when effects caused by the method of registration in a single-beam scheme are superimposed. For this, we develop a mathematical apparatus describing such a system under the condition of complete mixing of the Zeeman sublevels of the excited state. A number of simple models of the atom are considered and a comparison is made with the experimentally registered alignment signals in vapors of a real alkali metal (cesium). This allowed us to derive common patterns for them and to derive the basic rules linking the properties of LD symmetry with the properties of the anisotropy tensor of the medium.  Most importantly, we substantiate the fundamental possibility of the interesting effect of LD signal inversion in the single-beam scheme, previously described in
\cite{Petrenko_Pazgalev_Vershovskii_2024}. 

Here we must explain why this phenomenon is of interest. Light of different polarizations, interacting with different optical transitions, causes alignment of different signs in the atomic medium (positive alignment is usually considered to be alignment with predominant population of levels with maximum values of the magnetic moment modulus, and vice versa). Nevertheless, as will be shown below, when registering alignment by light of the same polarization on the same optical transition, the linear dichroism signals always have the same sign. From our calculations, however, it follows that this rule can actually be violated in systems characterized by overlapping of optical profiles due to different pumping and detection conditions on them. 

This effect is of undoubtable practical interest, since the spectral position of the LD signal zeroing points is characterized by high stability, and therefore can be used to stabilize the laser frequency. This is the basis for the TL-DAVLL (Transverse Linear Dichroic Atomic Vapor Laser Lock) method proposed in 
\cite{Petrenko_Pazgalev_Vershovskii_2024}. The method is based on switching the direction of a weak ($\sim$ 1~$\mu$T) magnetic field in a cell with alkali atoms in a plane perpendicular to the linearly polarized pump beam and on synchronous detection of the intensity of this beam at the cell output. When scanning the frequency of laser radiation in the vicinity of resonances, the sign of the signal at the output of the synchronous detector changes to the opposite, which allows it to be used to close the tracking feedback. The fundamental advantage of this method over existing ones is that it is quite simple, and it does not require either laser frequency modulation or strong magnetic fields, and therefore can be used in the most precise experiments on spectroscopy, cooling and interference of alkali atoms.   

  This article is organized as follows: in Section \ref{sec:2} we present the basic concepts of the theory of linear dichroism, describe simple models of alignment in the ground state of a one-electron atom, and analyze the dependencies that follow from them. Next, we extend the results obtained to systems with a complex spectrum and show that, under conditions of incomplete spin mixing in excited states, an inversion of linear dichroism signals can occur in such systems. In Section \ref{sec:3} we describe the experiment. In Section \ref{sec:4} we present the experimental results and compare them with the predictions of the above theory. Section \ref{sec:5} contains a discussion of the results obtained.

\section{THEORY}\label{sec:2}
\subsection{Linear dichroism}\label{sec:a}

\begin{figure*}[!t]  
	\includegraphics[width=\linewidth]{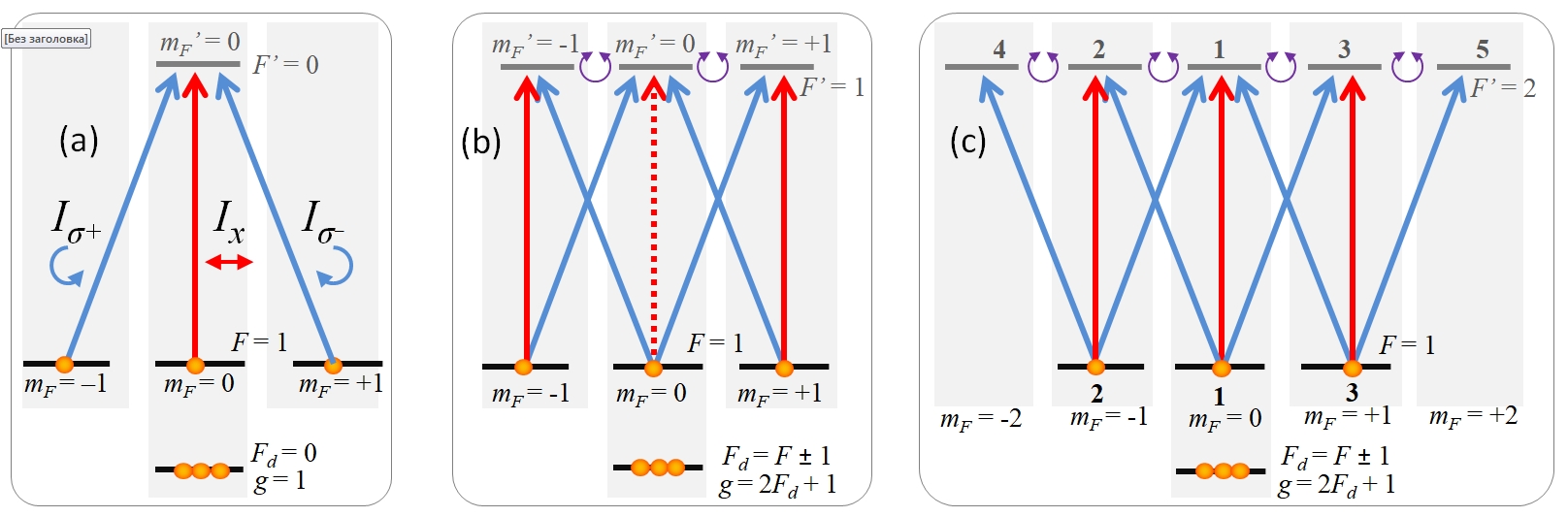}
	\caption{Models of systems with three ground state sublevels ($F = 1$) interacting with light: (a) $F' = 0$ (hydrogen), (b) $F' = 1$ (hydrogen,  \textsuperscript{87}Rb,  \textsuperscript{39}K,  \textsuperscript{41}K, the transition $m_F = 0 \leftrightarrow m_{F'} = 0$ is forbidden), (c) $F' = 2$ (\textsuperscript{87}Rb,  \textsuperscript{39}K,  \textsuperscript{41}K). The notations are the same as in  Figure \ref{fig1}. The numbering of sublevels in the model (c) is shown in bold; the numbering goes from the central levels to the outer ones. Levels $F_d = F \pm 1$ are “dark'', i.e. they do not interact with light.}\label{fig2}
\end{figure*}

In this section we briefly present the basic concepts and notations. Let the pump beam and the weak probe beam propagate along the $z$ axis, and the weak magnetic field be directed along the x axis. The LD in such a system is manifested in the difference in the absorption coefficients $K_x$ and $K_y$, measured by the probe beam of the corresponding polarization. The passage of the probe beam with the polarization azimuth $\varphi$ through a medium of length $L$ is described using the Jones formalism as follows:
\begin{equation}
  {E_{Lx'} \choose E_{Ly'}}
  =\mathbf{T}\cdot {E_{x'} \choose E_{y'}} 
  \label{eq:1}
\end{equation}
where $E_{x}=E \cdot Cos(\varphi)$, $E_{y}=E \cdot Sin(\varphi)$, $E$ is the amplitude of the electric component of the light wave, \textbf{T} is the Jones matrix containing real coefficients $T_{ij}$ (which corresponds to the absence of birefringence):
\begin{equation}
    \mathbf{T}={{T_{x'x'} \quad {0}} \choose {{0} \quad T_{y'y'}}} ,
    \label{eq:2}
\end{equation}
where the off-diagonal elements $T_{x'y'}$ and T$_{y'x'}$ are equal to zero 
(here and below we neglect the effects of the optically thick layer, considering $K_0 \ll 1$). In the simplest isotropic case, the bleaching of the medium by strong pumping light is described by the dependence
\begin{equation} 
K=\frac{K_{0} }{1+I/I_{{\rm sat}} }, 
\label{eq:3}
\end{equation}
where $K_0$ is the relative absorption in the medium in the absence of pump light, $I_{\text{sat}}$ is the saturation intensity of the transition. Note that Eq.~\eqref{eq:3}, unlike Eq.~\eqref{eq:1}, does not include the amplitudes of the electric component, but the intensities of the pump light. This is due to the fact that the rates of light-induced transitions from the ground to the excited state of the atom are proportional to the intensities of the light components.

In a system characterized by linear dichroism, the absorption coefficient can be described by a two-dimensional vector $\mathbf{K} = (K_x, K_y)^{\text{T}}$  (the index “T''  means transposition), and, as will be shown below, in the simplest case (in the approximation of weak pump intensity)
\begin{equation} 
\mathbf{K} = K_{0} (\mathbf{e}_{1} - \mathbf{a}\cdot \mathbf{I}),
\label{eq:4}
\end{equation}
where $\mathbf{e}_1 =~(1,1)^{\text{T}}$, $\mathbf{a}$ is the anisotropy tensor, $\mathbf{I}$ is the pump intensity vector:
\begin{equation}
\begin{array}{l} 
{\displaystyle \mathbf{a}={{a_{x,x} \quad a_{x,y}} \choose {a_{y,x} \quad a_{y,y}}},} \\ {}\\
{\displaystyle \mathbf{I}={I_{x} \choose I_{y}}={E_{x}^{*}  E_{x} \choose E_{y}^{*} E_{y}}} 
={\displaystyle {I_{0} \cos ^{2} (\varphi ) \choose I_{0} \sin ^{2} (\varphi )}}.
\end{array}
\label{eq:5}
\end{equation}

Expression \eqref{eq:4} assumes that in the absence of pumping the medium is isotropic, which is typical for atomic systems in the gas phase. LD in such systems arises as a result of the fact that the pump components, polarized parallel and perpendicular to the magnetic field, interact differently with the medium. LD of this kind can be called induced dichroism. Note that $\mathbf{K}$ is not a full-fledged vector: it can only be defined in a coordinate system rigidly related to the magnetic field. Let us define dichroism as the relative difference in absorption coefficients:
\begin{equation}
\begin{array}{cc} 
\Delta K\equiv \frac {\displaystyle K_{y} -K_{x} }{\displaystyle 2}, \\ \\
\bar{K}\equiv \frac{\displaystyle K_{y} +K_{x} }{\displaystyle 2}
\end{array}
\label{eq:6}
\end{equation}
($\Delta K \ll \bar{K}$). Let us consider a cell with alkali atoms and a buffer gas; the buffer gas pressure will be considered sufficient to ensure complete mixing of the excited states of the atom during its collisions with the atoms (or molecules) of the gas. In this case, optical pumping will cause depopulation of the Zeeman sublevels of the ground state, characterized by different rates; but the rates of re-population of all the ground state sublevels will be the same (the so-called depopulation pumping 
\cite{Happer_1972}). 

The level diagram of the Cs atom ($I = 7/2$) in the simplest case, namely, with pumping and detection by linearly polarized light on the $F = 4 \leftrightarrow F' = 3$ transition, is shown in Figure \ref{fig1}. The vertical arrows represent the linearly polarized light component (i.e. the pump component polarized along the $x$ axis). The inclined arrows represent left- and right-polarized light components, into which the pump component polarized along the $y$ axis is decomposed ($I_{\sigma +} = I_{\sigma -} = I_y/2$). The transition probabilities are specified by the Clebsch-Gordan coefficients 
\cite{Sobelman_2012,
Varshalovich_Moskalev_Khersonskii_1988}. Double arc-shaped arrows denote the mixing of excited states. The diagram includes 16 levels of the ground state. The pumping diagrams for the three remaining transitions of the D\textsubscript{1} line of Cs look similar.

If we include other one-electron atoms in our consideration, we will see that as the nuclear moment $I$ decreases, the level diagram simplifies. The level diagrams of $^{85}$Rb ($I = 5/2$) and $^{87}$Rb ($I = 3/2$) look similar to the diagram in Figure \ref{fig1}, but with a smaller number of levels.

Since $i$) the excited-state energy of such atoms in the entire range of operating temperatures of quantum sensors is many times greater than the thermal energy, and $ii$) the relaxation rate of the excited state, as a rule, exceeds the relaxation rates of the ground state sublevels by 3-5 orders of magnitude, the populations of the excited states can be considered equal to zero. In the absence of pumping, we will consider the populations of the ground state sublevels to be uniformly populated -- this is true until the Zeeman splitting becomes comparable to the thermal energy. Assuming the absence of any coherences in the ground state (which is valid for a sufficiently strong magnetic field and stationary experimental conditions) and complete mixing of the excited state levels, the population distribution in such a system can be calculated by solving a system of balance equations (see Appendix~A). 

The simplest model in which alignment is possible and by means of which it is possible to describe the LD effects under pumping with linearly polarized light should include three Zeeman sublevels in the lower state Figure \ref{fig2}. Such a model adequately describes the hydrogen atom H ($I = 1/2$) on the transition $F=1 \leftrightarrow F'= 0$ (Figure \ref{fig2}a), the hydrogen atom H on the transition $F=1 \leftrightarrow F'= 1$ (Figure \ref{fig2}b), the atoms of rubidium $^{87}$Rb, potassium $^{39}$K and $^{41}$K and other alkali metals with $I = 3/2$ on the transition $F = 1 \leftrightarrow F' =2$ (Figure \ref{fig2}c).

The presence of the second hyperfine “dark'' level in all these atoms leads to a general decrease in absorption and, as will be shown below, also affects the dichroism effects. In the model, the “dark'' Zeeman sublevels are considered as one level with statistical weight $g = 2F_d + 1$, where $F_d$ is the total moment of the dark state. Note that for more complex atomic systems it is necessary to calculate the effective statistical weight taking into account the total number of levels. In the scheme of Figure \ref{fig2}b there are three excited sublevels, and in the scheme of Figure \ref{fig2}c there are five, but mixing during collisions with the buffer gas equalizes the populations on them. The transition probabilities in the schemes shown in Figure \ref{fig2} are given in Table~\ref{tab:t1} (Appendix~A). 

In the case of linear polarization of the pump light (due to the symmetry of the conditions) in these models the populations of the extreme levels of the ground state $m_F = -1$ and $m_F = +1$ are equal. The probabilities of symmetric transitions are also equal.

Solutions of the balance equations for these systems are given in Appendix~A. In them, we can identify the terms responsible for the constant absorption $\mathbf{p}_0 =(p_{x0}, p_{y0})^{\text{T}}$ and for the induced linear dichroism $\mathbf{\Delta} = (\Delta_x, \Delta_y)^{\text{T}}$:
\begin{equation}
\mathbf{p}={{p_{x0} +2\Delta _{x} \quad p_{y0} +2\Delta _{y}} \choose {p_{x0} -\Delta _{x} \quad p_{y0} -\Delta _{y}}} .
\label{eq:7}
\end{equation}

With some degree of conventionality, we can say that the vector $\mathbf{\Delta}$ characterizes the “linear polarizability'' of the medium, and its components $\Delta_x$, $\Delta_y$ characterize the longitudinal and transverse “linear polarizability'', respectively.

As will be seen below, it is precisely this choice of normalization of $\Delta $ that provides the most simple and symmetric form of the equations. Parameters $\Delta_x$, $\Delta_y$ will be further referred to as induced dichroism coefficients.

Assuming $p_{x0}=p_{y0}=1$ (Appendix~A), we see that the solution of system Eqs.~\eqref{eq:A3}, \eqref{eq:A6} in general form looks like this:
\begin{equation}
\begin{array}{c} 
{K_{x} =K_{0}  \left(1-\frac{\displaystyle
a_{x,x} I_{x} +a_{x,y} I_{_{y} } +c_{0} (I_{x} ,I_{y} )}{\displaystyle
1+b_{x,x} I_{x} +b_{x,y} I_{y} +c_{0} (I_{x} ,I_{y} )} \right),} \\ {}\\
{K_{y} =K_{0} \left(1-\frac{\displaystyle
a_{y,x} I_{x} +a_{y,y} I_{_{y} } +c_{0} (I_{x} ,I_{y} )}{\displaystyle
1+b_{y,x} I_{x} +b_{y,y} I_{y} +c_{0} (I_{x} ,I_{y} )} \right),}
\end{array}
\label{eq:8}
\end{equation}
where 
\begin{equation}
\begin{array}{l} {\mathbf{a}\equiv \left(\begin{array}{cc} {a_{x,x} } & {a_{x,y} } \\ {a_{y,x} } & {a_{y,y} } \end{array}\right)} \\{}\\ {\qquad =q\left(\begin{array}{cc} {1} & {1} \\ {1} & {1} \end{array}\right) +2\left(\begin{array}{cc} {\Delta _{x} \Delta _{x} } & {\Delta _{x} \Delta _{y} } \\ {\Delta _{x} \Delta _{y} } & {\Delta _{y} \Delta _{y} } \end{array}\right),} \\{}\\ 
{\mathbf{b}\equiv \left(\begin{array}{cc} {b_{x,x} } & {b_{x,y} } \\ {b_{y,x} } & {b_{y,y} } \end{array}\right)} \\{}\\ {\qquad=(1+q)\left(\begin{array}{cc} {1} & {1} \\ {1} & {1} \end{array}\right) +\left(\begin{array}{cc} {\Delta _{x} } & {\Delta _{y} } \\ {\Delta _{x} } & {\Delta _{y} } \end{array}\right)}, \end{array}
\label{eq:9}
\end{equation}
$q = g/(3+g)$. The two rows of the matrix $\mathbf{b}$ are the same, so we can replace it with a vector $\mathbf{b} = (b_x, b_y)^{\text{T}}$. Eq.~\eqref{eq:9} can be written in a compact form:
\begin{equation}
\begin{array}{l} {\mathbf{a}=q\mathbf{e}_{1}^{\text{T}} \otimes \mathbf{e}_{1} +2\mathbf{\Delta}^{\text{T}} \otimes \mathbf{\Delta},} \\{}\\ {\mathbf{b} = (1+q)\mathbf{{e}_{1} +\Delta ^{\text{T}}.}} 
\end{array}
\label{eq:10}
\end{equation}

The term $c_0(I_x, I_y)$ contains terms quadratic in intensity:
\begin{equation}
\begin{array}{l} {c_{0} (I_{x} ,I_{y} )  = q((1+2\Delta _{x} )I_{x}}\\{}\\ { +(1+2\Delta _{y} )I_{y} )((1-\Delta _{x} )I_{x}+(1-\Delta _{y} )I_{y} ).} 
\end{array}
\label{eq:11}
\end{equation}
Thus, taking into account the “dark'' levels leads to the appearance of a term proportional to the square of the pump intensity. At $g = 0$ (“dark'' levels are absent), Eqs.~\eqref{eq:8} take the form Eq.~\eqref{eq:4}.

Note that the two-dimensional matrix a is symmetric with respect to the main diagonal. Substituting Eq.~\eqref{eq:5} into \eqref{eq:A5} and introducing the following notations:
\begin{equation}
\begin{array}{l} {\Delta _{\pm } =\frac{\displaystyle \Delta _{x} \pm \Delta _{y} }{\displaystyle 2} ,} \\{}\\ {v=\Delta _{+} +\Delta _{-} \cos ({\rm 2}\varphi ),} \\{}\\  {w=q(1-v)(1+2v),} \\{}\\  {{\rm \beta }=1+q+v+I_{0} w} ,
\end{array}
\label{eq:12}
\end{equation}
where $\varphi$ is the angle between the magnetic field and the plane of polarization of light, we obtain 
\begin{equation}
\begin{array}{l} {\bar{K}=K_{0} \left(1-I_{0} \frac{\displaystyle q+2\Delta _{+} v+I_{0} w}{\displaystyle 1+{\rm \beta }I_{0} } \right),} \\{}\\  {\Delta K=K_{0} v\frac{\displaystyle 2 I_{0} \Delta_{-}  }{\displaystyle 1+{\rm \beta }I_{0} } ,} 
\end{array}
\label{eq:13}
\end{equation}
and for the alignment multipole
\begin{equation} 
{\rm \rho }_{{\rm 0}}^{{(2)}} =\sqrt{\frac{2}{3} } (1-q)v\frac{I_{0} }{1+{\rm \beta }I_{0} }.
\label{eq:14}
\end{equation}

Thus, all angular dependencies of the absorption coefficients in the case of alignment are reduced to combinations of even angular dependencies $v$. At low intensities
\begin{equation} 
{\rm \rho }_{{\rm 0}}^{{(2)}} \approx \sqrt{\frac{2}{3} } (1-q)I_{0} (\Delta _{+} +\Delta _{-} \cos (2\varphi )).
\label{eq:15}
\end{equation}

Thus, the coefficient $\Delta_+$ determines the uniform contribution to the alignment over the angle, and $\Delta_-$ determines the contribution of the angle-dependent component. As we can see, this component is proportional to $\cos(2\varphi)$; at low pump intensities the alignment multipole does not contain other angular dependencies.

Note that due to the universality of the concept of the alignment multipole, all the calculations obtained for a system with $F = 1$ and presented in this section, can be used with varying degrees of accuracy for systems with a larger number of levels. In this case, the coefficients $i = c, e$ of the matrix $\mathbf{p}$ will correspond to the groups of levels (with the corresponding weights), giving negative and positive contributions to $\rho^{(2)}_0$, respectively. For $F = 2$, these are, respectively, the levels $m_F=0,\pm1$ and $m_F=\pm2$.

\begin{figure*}[!t]  
	\includegraphics[width=\linewidth]{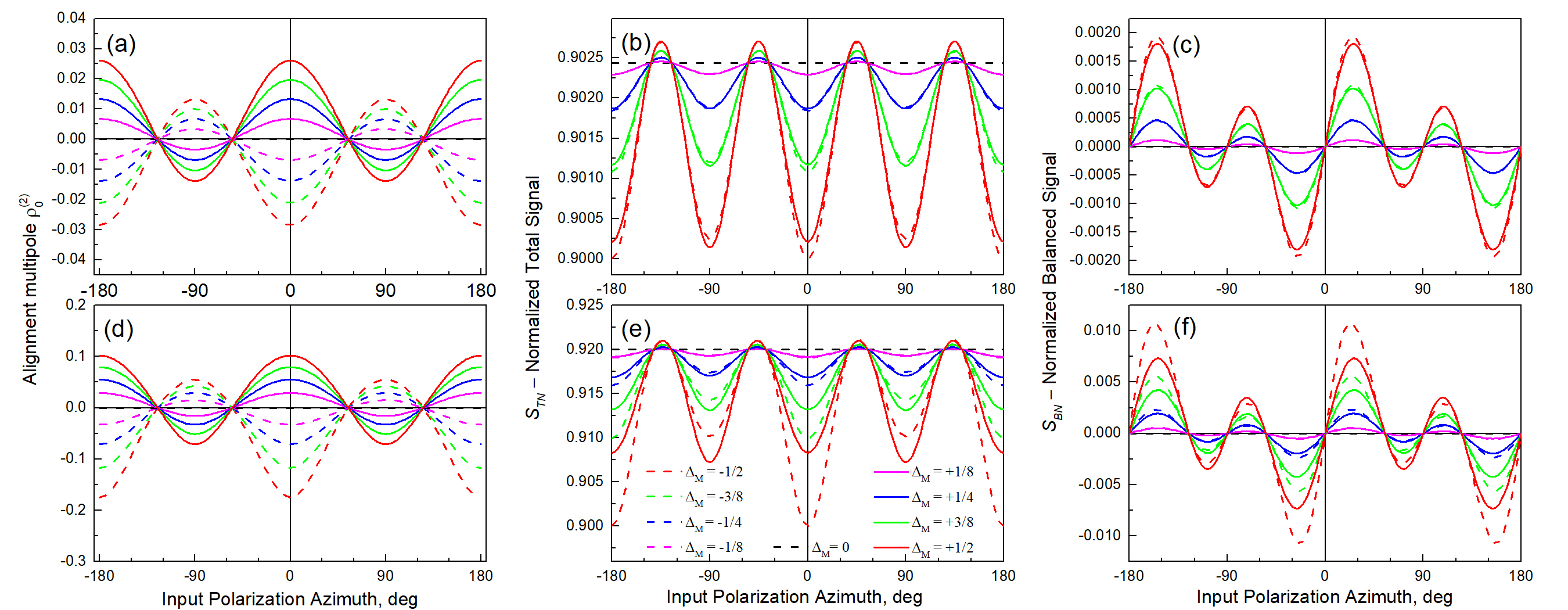}
	\caption{Calculated angular dependences (a), (d) of the zero alignment multipole $\rho^{(2)}_{0}$, (b), (e) normalized total signal  $S_{TN}$, (c), (f) normalized balanced signal $S_{BN}$ in a single-beam scheme at $I_0 = 0.1$ ((a), (b), (c)) and $I_0 = 1$ ((d), (e), (f)) for different values of $\Delta_{\tt{M}}$.}\label{fig3}
\end{figure*}

\subsection{Single-beam scheme}\label{sec:b}

Let us proceed to consider a particular case, but one that is of maximum interest from a practical point of view -- a single-beam scheme. In this scheme, the pumping light simultaneously performs the function of probe light. In this case, the previously derived angular dependencies, caused by the rotation of the pump polarization azimuth, are superimposed on the dependence caused by the rotation of the probe light polarization. Substituting the light parameters $I_0$, $\varphi$ into Eqs.~\eqref{eq:1}--\eqref{eq:5}, we obtain for a thin layer: 
\begin{equation}
\begin{array}{l} {K=\bar{K}-\Delta K\cos (2\varphi )} \\{}\\  {\qquad =K_{0} (1-\frac{\displaystyle 1}{\displaystyle 1+{\rm \beta }I_{0} } (q+2v^{2} +I_{0} w))} \\{}\\ 
{\qquad =\frac{\displaystyle K_{0} }{\displaystyle 1+{\rm \beta }I_{0} } (1+I_{0} \frac{\displaystyle w}{\displaystyle q} ).} \end{array}
\label{eq:16}
\end{equation}

In the limit $I_0 \ll ~1$, the angular dependence can be excluded from the denominator using the Taylor series expansion method:
\begin{equation}
\begin{array}
{l} {K\approx \frac{\displaystyle K_{0} }{\displaystyle 1+2bI_{0} +b^{2} I_{0}^{2} } \times (1+I_{0} (2b-a)} \\{}\\ 
{\qquad -I_{0} \Delta _{-}^{2} (1+4\frac{\displaystyle \Delta_{+} }{\displaystyle \Delta_{-} } \cos (2\varphi )+\cos (4\varphi ))+{\varepsilon }I_{0}^{2}),} 
\end{array}
\label{eq:17}
\end{equation}
where  
\begin{equation}
{\rm \varepsilon }=\frac{w}{q} (1+\Delta _{+} -\Delta _{-} \cos (2\varphi )), 
\label{eq:18}
\end{equation}
and $a$, $b$ are the average values of the matrices $\mathbf{a}$ and $\mathbf{b}$:
\begin{equation}
\begin{array}{l} {a=q+2\Delta _{+}^{2}, } \\{}\\  {b=1+q+\Delta _{+} } .
\end{array}
\label{eq:19}
\end{equation}

Eq.~\eqref{eq:17} can be further simplified by discarding the terms proportional to $I_0^2$. It can be seen that Eq.~\eqref{eq:17} contains terms responsible for both isotropic and anisotropic bleaching of the medium, and the latter can be decomposed into two components with a “dipole'' and “quadrupole'' (in terms of 
\cite{Fomin_Kozlov_Petrov_Smirnov_Petrenko_Zapasskii_2025}
)) angular dependence even at $I_0 \rightarrow 0$.

As shown in 
\cite{Budker_Gawlik_Kimball_Rochester_Yashchuk_Weis_2002,
Fomin_Kozlov_Petrov_Smirnov_Petrenko_Zapasskii_2025,
Meraki_Elson_Ho_Akbar_Kozbial_Kolodynski_Jensen_2023}, in a single-beam scheme it is possible to measure LD by measuring either the total intensity of the transmitted light or the rotation of its polarization due to anisotropic absorption. Hereinafter we will call the corresponding signals $S_T$ (total) and $S_B$ (balanced). The easiest way to measure $S_B$ is to split the beam into two perpendicularly polarized components with initially equal intensities. The axis of the polarization-separating element is set at an angle of $\pi$/4 to the initial azimuth of the beam polarization, the intensity of the components is measured by two photodetectors, the signals of which are subtracted to measure the $S_B$ signal, and can be summed to measure the $S_T$ signal. Expressions linking signal values with system parameters are given in Appendix~B.

Since, as noted earlier, all absorption coefficients for linearly polarized pumping are even functions of angle, it can be seen that the $S_B$ signal is an odd function of angle, passing through zero values at polarization azimuths that are multiples of $\pi $/4. This property of the balanced signal is an additional advantage that makes it attractive for a number of applications.

\subsection{The “magic'' angle}\label{sec:c}

Let us consider one distinctive feature of systems with alignment: the presence of a “magic'' angle, that is, the angle between the plane of polarization of light and the field at which alignment is absent 
\cite{Budker_Kimball_DeMille_2004}. In a system with complete mixing of the excited state sublevels the “magic'' angle $\varphi_{\text{M}} = \arccos(1/\sqrt{3})$, which corresponds to the condition $I_y = 2I_x$ (Eq.~\eqref{eq:5}). The presence of a “magic'' angle is due to the general symmetry properties of the system -- namely, as stated above, the component $I_x$, polarized along the magnetic field, always causes transitions of one type ($\Delta m_F=0$), and the component $I_y$, polarized perpendicular to the field, is decomposed into two circularly polarized components, causing transitions $\Delta m_F=\pm1$; each of these components has half the intensity of the $I_y$ component. According to 
\cite{Budker_Kimball_DeMille_2004}, the angular dependence of alignment is described by the second Legendre polynomial
\begin{equation}
P_{L2} =-\frac{1}{2} (1-3\cos ^{2} (\varphi ))=\frac{1}{4} (1+3\cos (2\varphi )).
\label{eq:20}
\end{equation}

The “magic'' angle condition is satisfied if the following values are substituted into Eq.~\eqref{eq:12}, \eqref{eq:13}:
\begin{equation}
\begin{array}{l}
\Delta_{x} = -2\Delta_{y}, \\{}\\  
\Delta_{-} =3 \Delta_{+}. 
\end{array}
\label{eq:21}
\end{equation}

It is easy to verify that the calculated values of the absorption probabilities for all transitions of all alkali metals satisfy condition Eq.~ \eqref{eq:21} -- see, for example, Table~\ref{tab:t2} (Appendix~B), which presents the coefficients of the optical pumping rates $p_{i,j}$ for the magnetic sublevels of the ground state of atoms with $F = 1$, and the corresponding values $\Delta_x, \Delta_y, \Delta_{-}, \Delta_+$. Thus, we now have only one independent coefficient describing the linear polarizability of the system
\begin{equation}
\Delta_{{\rm M}} =\Delta_{x} =-2\Delta_{y} = 4\Delta_{+} =\frac{4}{3} \Delta_{-} 
\label{eq:22}
\end{equation}

The transition probability matrix $\mathbf{p}$ takes the form
\begin{equation}
\mathbf{p}=\left(\begin{array}{cc} {1+2\Delta _{{\rm M}} } & {1+\Delta _{{\rm M}} } \\ 
{1-\Delta _{{\rm M}} } & {1+\frac{\displaystyle \Delta _{{\rm M}} }{\displaystyle 2} } \end{array}\right)
\label{eq:23}
\end{equation}

Linear dichroism Eq.~\eqref{eq:6}, according to Eq.~\eqref{eq:8} and Eq.~\eqref{eq:C1} (see Appendix~C for details), is expressed through the alignment multipole as follows:
\begin{equation}
\Delta K=\sqrt{\frac{3}{2} } \frac{3}{1-q} \Delta _{{\rm M}} {\rho }_{0}^{(2)}, 
\label{eq:24}
\end{equation}
and for $I_0 \ll 1$
\begin{equation} 
{\rho }_{0}^{(2)} =\sqrt{\frac{2}{3} } (1-q)I_{0} \Delta _{{\rm M}} P_{L2}
\label{eq:25}
\end{equation}
and
\begin{equation}
\Delta K=\frac{3}{2} I_{0} {\Delta }_{{\rm M}}^{{2}} P_{L2}.
\label{eq:26}
\end{equation}

\begin{figure}[!t]  
	\includegraphics[width=\linewidth]{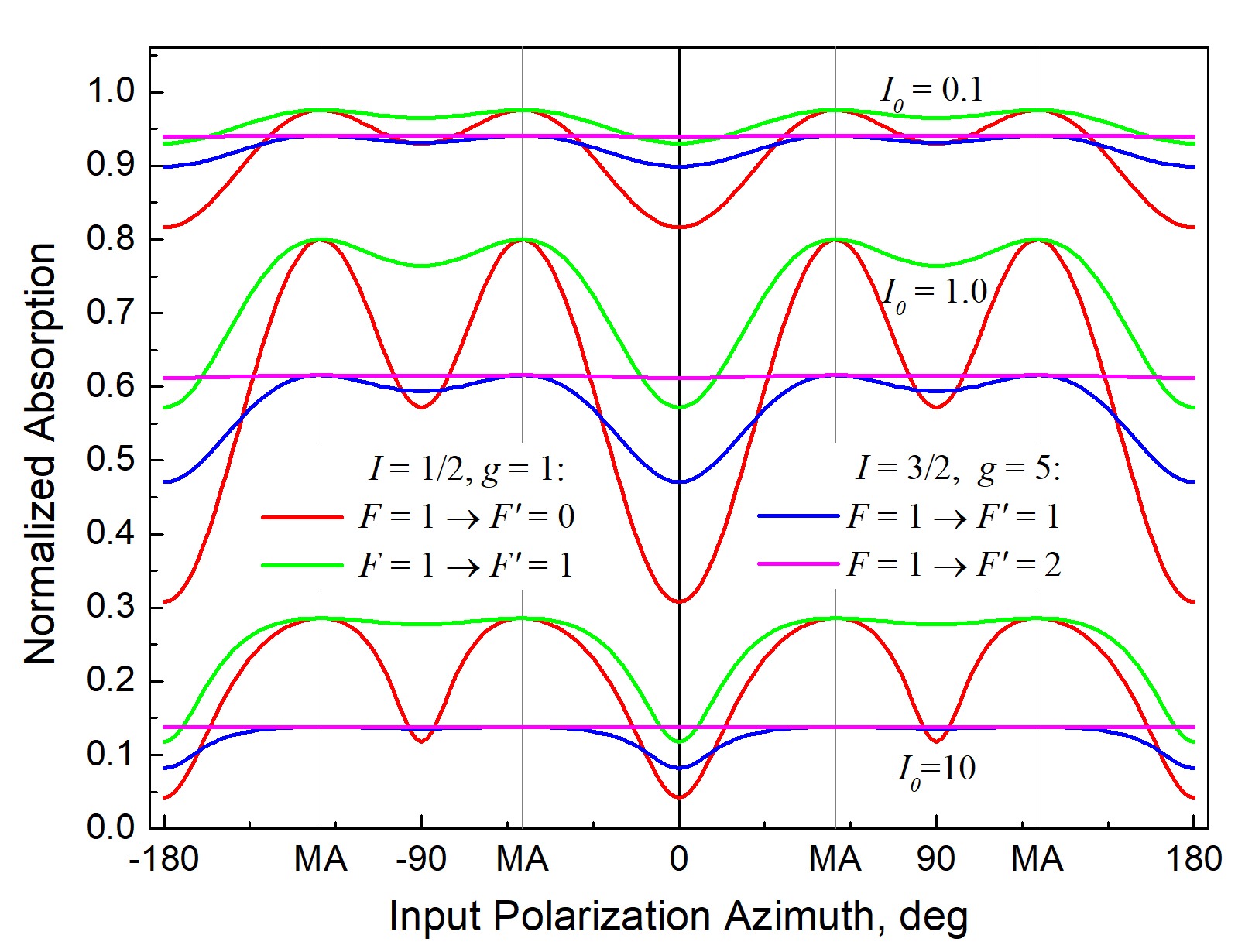}
	\caption{Angular dependences of the normalized absorption coefficient $K/K_0$ at $I_0 = 0.1, 1, 10$ in a single-beam scheme with optical pumping from the level $F = 1$ in different schemes (see Figure \ref{fig3}).}\label{fig4}
\end{figure}

As expected, in the weak pump approximation, the angular dependence of the LD is described by the second Legendre polynomial of the cosine of the angle $\varphi$ -- the function $P_{L2}(\varphi)$, which in terms of work 
\cite{Fomin_Kozlov_Petrov_Smirnov_Petrenko_Zapasskii_2025} corresponds to a purely “dipole'' character of the angular dependence of the LD. Higher even harmonics of the angle $\varphi$ appear in Eq.~\eqref{eq:24} as the pump intensity increases. Figure \ref{fig3} shows the calculated angular dependencies of the zero alignment multipole $\rho^{(2)}_0$ and the LD signals in a single-beam scheme for $I_0=0.1$ and $I_0~=~1$ depending on the value of $\Delta_\text{M}$. It is evident that the alignment of opposite signs leads to similar values of the LD signals, since the medium is interrogated by the same light that causes pumping. Correspondingly, LD signal inversion is not observed under any conditions.

Figure \ref{fig4} shows the angular dependencies of the absorption coefficient $K/K_0$ for three values of $I_0$ in a single-beam scheme with optical pumping from the level $F=1$ in different schemes (see Figure \ref{fig2}).

It should be noted that the dichroism value strongly depends on the statistical weight of the “dark'' level. Next we will show that if the condition for complete mixing of excited states is not met (for example, at low buffer gas pressure, or due to partial conservation of nuclear spin during collisions in the excited state 
\cite{Popov_Bobrikova_Voskoboinikov_Barantsev_Ustinov_Litvinov_Vershovskii_Dmitriev_Kartoshkin_Pazgalev_2018}), then the probabilities of relaxation from the excited state to the ground state are not equal, and the conclusions of this section may be violated.

\subsection{Alignment in systems with complex spectrum}\label{sec:d}
As shown in \cite{Petrenko_Pazgalev_Vershovskii_2024}, in a real single-beam scheme, in the presence of closely spaced transitions with overlapping optical profiles, the linear dichroism signal can be inverted, and the ``magic''  angle condition may be violated. Let us assume that in the optical spectrum, the atomic optical profiles partially overlap in pairs -- as happens in the spectrum of the D\textsubscript{1} line of alkali metals, when the broadening by collisions with atoms (molecules) of the buffer gas becomes comparable to the value of the hyperfine splitting of the excited state. Let us assume that the dichroism parameters of two adjacent transitions satisfy condition Eq.~\eqref{eq:21} and are characterized by the coefficients $\Delta_{\rm M1}$ and $\Delta_{\rm M2}$, respectively; the transitions are affected by the common pump light, and the spectral intensities of the pump at these transitions are equal to 
\begin{equation} 
I_{0i} =I_{0} f_{i} L(\nu -\nu_{i} ) ,
\label{eq:27}
\end{equation}
where
\begin{equation}
L(x)\equiv \frac{x^{2} }{x^{2} +\Gamma ^{2} }. 
\label{eq:28}
\end{equation}
is the Lorentz contour, $f_i$ is the relative transition amplitude (oscillator strength), $\nu - \nu_i$ is the frequency detuning of light from the corresponding transition, and $\Gamma$ is the width (HWHM) of the optical transition, which is considered the same for all levels of the hyperfine structure. It is assumed that $\Gamma$ significantly exceeds the value of the Doppler broadening -- this is true at gas pressures exceeding several tens of Torr. Then \eqref{eq:24} can be rewritten as
\begin{equation}
\Delta K=\sqrt{\frac{3}{2} } \frac{3}{1-q} \sum _{i,j}\Delta _{{\rm M}j} {\rm \rho }_{0i}^{2},
\label{eq:29}
\end{equation}
where index $i = 1, 2$ corresponds to pumping (polarized at an angle $\varphi$ to the field), $j = 1, 2$ -- to detection. Substituting expression \eqref{eq:26} into \eqref{eq:29}, for $I_0 \ll 1$ we obtain
\begin{equation}
\Delta K({\rm \nu ,}\varphi )=\frac{3}{2} I_{0} P_{L2} \sum _{i,j}f_{i} L({\rm \nu }-{\rm \nu }_{i} )\Delta _{{\rm M}i} \Delta _{{\rm M}j} .
\label{eq:30}
\end{equation}

The difference in the intensities of light absorbed along the $x$ and $y$ axes in a single-beam scheme ($\varphi =0, \pi /2$) will be equal to a first approximation
\begin{equation}
\begin{array}{l} {\Delta I({\rm \nu })\approx \Delta K({\rm \nu })I_{d} ({\rm \nu })} \\{}\\ {=-AI_{0}^{2} (f_{1} \Delta _{{\rm M}1} L({\rm \nu }-{\rm \nu }_{1} )-f_{2} \Delta _{{\rm M}2} L({\rm \nu }-{\rm \nu }_{2} ))^{2} ,} 
\end{array}
\label{eq:31}
\end{equation}
where $A > 0$ is the normalization coefficient. The values of $\Delta_{\rm M1}$, $\Delta_{\rm M2}$ for two optical transitions from the same hyperfine level of the ground state always have different signs (see Table~\ref{tab:t2}). But, as follows from Eq.~\eqref{eq:31}, direct transitions in a single-beam scheme always cause LD signals of the same sign; they vanish at the light frequency at which the probabilities of the $1 \leftrightarrow 1$ and $2 \leftrightarrow 2$ transitions are equal.

\subsection{LD signals inversion}\label{sec:e}

Contrary to the conclusions of the previous section, in our previous work 
\cite{Petrenko_Pazgalev_Vershovskii_2024}
 it was shown that under certain conditions the LD signals can change sign; this occurs precisely in the region of overlap of the contours $L(\nu - \nu_1)$ and $L(\nu - \nu_2)$ and the maximum of the inverted LD coincides with the maximum of the overlap. It follows from this that expression Eq.~\eqref{eq:31} needs to be updated: we must admit that the efficiencies of interaction of the atom with light can differ for the pumping and detection processes. Thus, due to the previously mentioned phenomenon of partial conservation of nuclear spin in collisions in the excited state 
\cite{Popov_Bobrikova_Voskoboinikov_Barantsev_Ustinov_Litvinov_Vershovskii_Dmitriev_Kartoshkin_Pazgalev_2018}, the pumping efficiency decreases, while this phenomenon does not affect the detection efficiency.

Let us introduce into Eqs.~\eqref{eq:25}, \eqref{eq:30} corrections  $k_i \approx 1$ to the oscillator strengths only for pumping processes: 
\begin{equation} 
I_{0i} = I_{0} k_{i} f_{i} L({\nu }-{\nu }_{i} ).
\label{eq:32}
\end{equation}

The value $1 -k_i$ has the physical meaning of a relative decrease in the pumping rate due to the conservation of the nuclear spin. It can be roughly estimated using the method presented in 
\cite{Popov_Bobrikova_Voskoboinikov_Barantsev_Ustinov_Litvinov_Vershovskii_Dmitriev_Kartoshkin_Pazgalev_2018}, if we assume that for nuclear spin $I \gg s$, where $s=1/2$ is the electron spin,  $k_i \approx 1 -\alpha$, where $\alpha$ is the fraction of collisions in which the nuclear spin projection $m_I$ is conserved; accordingly, $k_i$ can be defined as the “mixing degree'' of nuclear spin. A more precise consideration should take into account that even with complete conservation of nuclear spin ($\alpha  = 1$), the pumping efficiency $k_i$ does not fall to zero, but to a value of $s/F$.  It should also be taken into account that the overall mixing degree of Zeeman sublevels in the excited state is determined by the ratio of the collision frequency to the lifetime of the excited state ($\sim$32~ns for Cs), and therefore depends on the gas pressure, but the mixing degree of the nuclear spin is determined by the ratio of the collision time to the inverse frequency of hyperfine splitting 
\cite{Popov_Bobrikova_Voskoboinikov_Barantsev_Ustinov_Litvinov_Vershovskii_Dmitriev_Kartoshkin_Pazgalev_2018}, and does not depend on the gas pressure. It is assumed that the introduced correction is the same for all Zeeman transitions within the hyperfine multiplet, and the condition of the “magic'' angle is not substantially violated for every single transition. Then
\begin{equation}
\begin{array}{l}
\Delta I({\nu }) =-BI_{0}^{2} \\{}\\
\begin{array}{l} {\times((r_{1} f_{1} \Delta_{{\rm M}1} L({\nu }-{\rm \nu }_{1} )-r_{2} f_{2} \Delta _{{\rm M}2} L({\rm \nu }-{\nu }_{2} ))^{2} } \\{}\\{-(r_{1} -r_{2} )^{2} f_{1} f_{2} |\Delta _{{\rm M}1} \Delta _{{\rm M}2} |L({\nu }-{\nu }_{1} )L({\nu }-{\nu }_{2} )),} \end{array} 
\end{array}
\label{eq:33}
\end{equation}
where
\begin{equation}
\begin{array}{l} {B=A (\frac{\displaystyle k_{1} +k_{2} }{\displaystyle 2} ),} \\ {} \\ {r_{i} =\sqrt{\frac{\displaystyle 2k_{i} }{\displaystyle k_{1} +k_{2} } } .} 
\end{array}
\label{eq:34}
\end{equation}
Now, for $r_1 \neq r_2$ (or, equivalently, $k_1 \neq k_2$), the value of $\Delta I(\nu )$ changes sign in the region of overlap of the contours $L(\nu - \nu_1)$ and $L(\nu - \nu_2)$. As follows from Eq.~\eqref{eq:33}, the LD inversion becomes possible due to “cross'' detection, when the alignment created at the transition $F=I+1/2 \leftrightarrow F'=I+1/2$ is detected at the transition $F=I+1/2 \leftrightarrow F'=I-1/2$ and vice versa. The results of the comparison of the model calculation using formula \eqref{eq:33} with the experimental data will be shown below. 

As mentioned in the Introduction, this result is of significant practical interest since it provides a theoretical basis for the LD inversion effect, which in turn underlies the principle of our recently proposed TL-DAVLL laser light stabilization method 
\cite{Petrenko_Pazgalev_Vershovskii_2024}.

\begin{figure*}[!t]  
	\includegraphics[width=\linewidth]{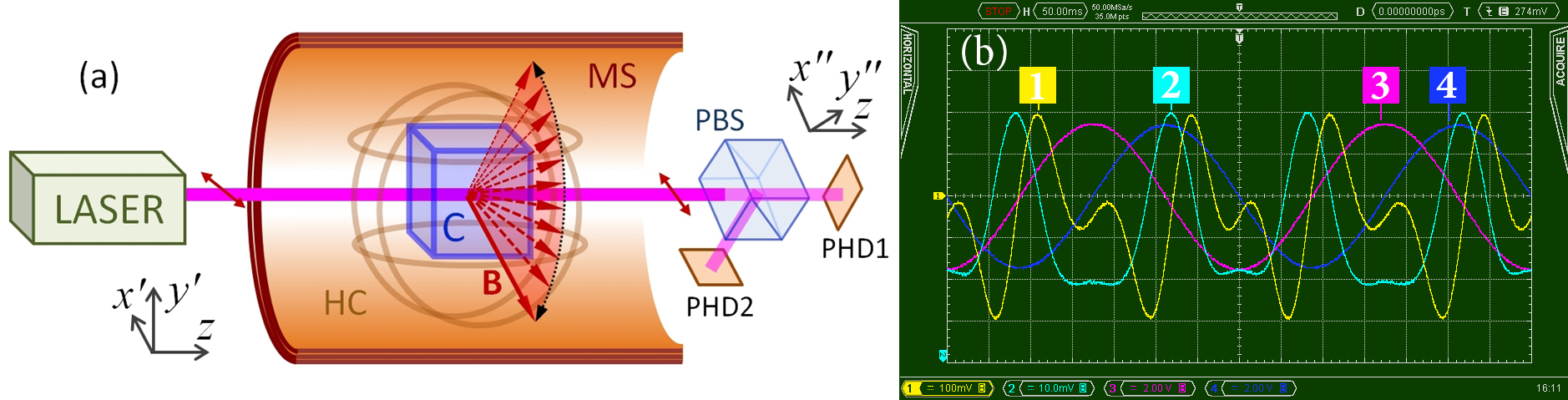}
	\caption{(a) Block diagram of the setup: MS -- magnetic shield, HC -- Helmholtz coil system, C -- cell, PBS -- polarizing beamsplitting cube, PHD1, PHD2 – photodiodes, $\mathbf{B}$ -- magnetic field vector. The magnetic field in the coils rotates in the $0x'y'$ plane; (b) Example of an oscillogram: 1 -- differential signal $S_B$, 2 -- total signal $S_T$, 3 -- current in the Y-coils, 4 -- current in the X-coils.}\label{fig5}
\end{figure*}

\section{EXPERIMENTAL SETUP}\label{sec:3}

A scheme of the experimental setup is shown in Figure \ref{fig5}a. A cell with an internal size of 5$\times$5$\times$5 mm$^3$ containing a few milligrams of Cs and nitrogen at a pressure of 200~Torr was placed in a thermostat (not shown in the scheme) and in a cylindrical three-layer magnetic shield. A VitaWave external-cavity diode laser tuned to the Cs D\textsubscript{1} line (wavelength 895~nm) was used as a pump/detection source. The laser frequency was scanned in the vicinity of the absorption lines $F=4 \leftrightarrow F'=3, 4$ (low-frequency pair) or $F~=~3 \leftrightarrow F'~=~3, 4$  (high-frequency pair). The range of continuous frequency tuning of the laser radiation was about 15~GHz.  The magnetic resonance width in the studied cell was 260~Hz, or 74~nT.

The magnetic field strength ($\sim1~\mu$T) was chosen such that the Larmor frequency $\omega_L$ significantly exceeded the relaxation rate $\Gamma$ of the ground state of Cs in the cell. This provided conditions under which transverse alignment components (coherences) could be neglected. The polarization azimuth was fixed, and the magnetic field vector rotated in a plane perpendicular to the beam (Figure \ref{fig5}), with the rotation rate $\omega_R$ being much less than the relaxation rate of the ground state: $\omega_R \ll \Gamma \ll \omega L$. 

The total ($S_T$) and differential ($S_B$) absorption signals in the cell were studied at various laser beam detunings in the Voigt geometry as a function of the angle between the magnetic field vectors and the azimuth of the pump light polarization (the same light served as a probe light). A polarizing beamsplitting cube (PBS) was used to separate the two polarizations. The axis of the PBS was set at an angle of 45$^\circ$ to the initial laser beam polarization. The intensity of the beam components was measured by photodiodes, the signals of which were subtracted to measure the $S_B$ signal, and were summed to measure the $S_T$ signal. The signals were measured in the photodetector coordinate system $x'' y'' z$, then converted into the laboratory coordinate system $x'y'z$ and into the $xyz$ coordinate system associated with the magnetic field ($\mathbf{B} || \mathbf{x}$). A more detailed description of the setup is given in \cite{Petrenko_Pazgalev_Vershovskii_2024}.

\section{EXPERIMENTAL RESULTS}\label{sec:4}

\begin{figure*}[!t]  
	\includegraphics[width=\linewidth]{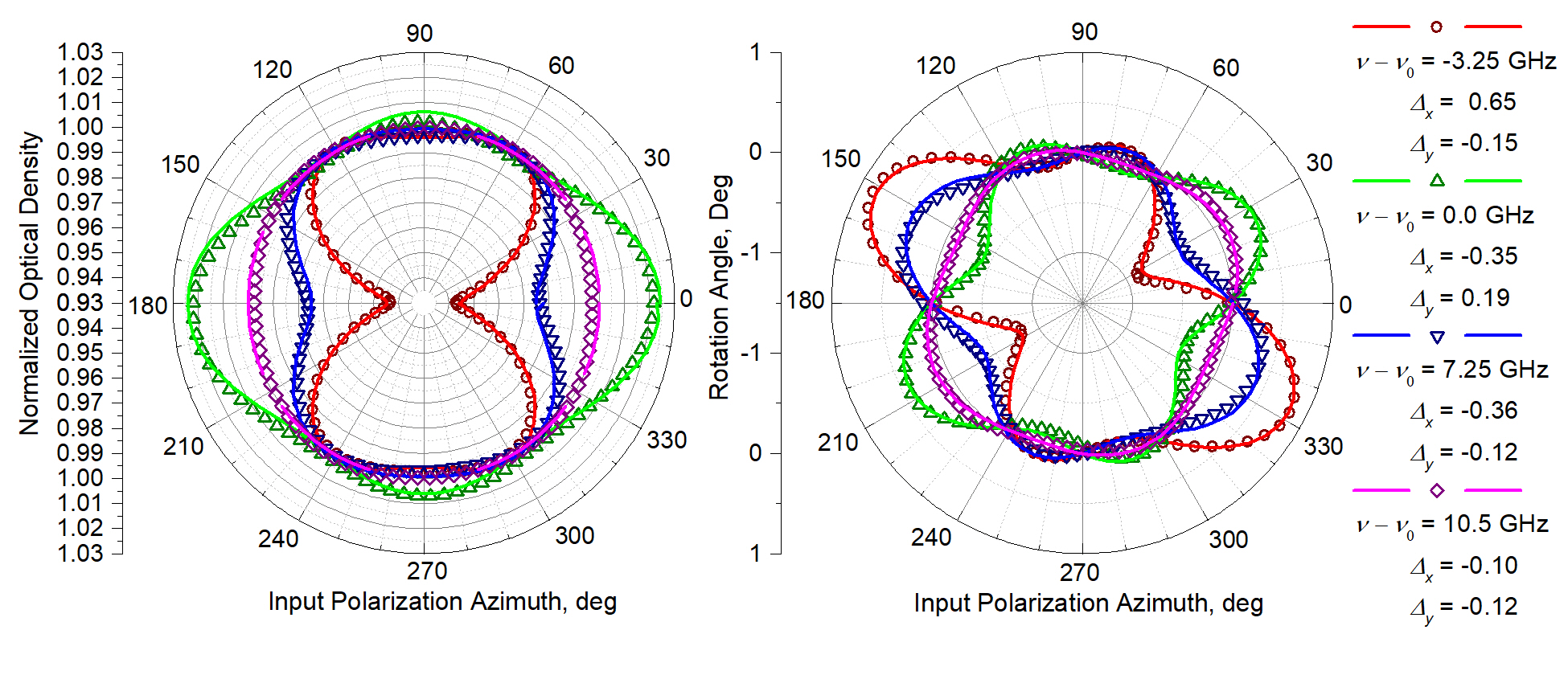}
	\caption{Examples of dependencies of the normalized optical density (left) and the rotation angle (right) on the azimuth of the input light polarization at different laser frequency detunings $\nu-\nu_0$ in the cell filled with 200 Torr N$_2$, $\nu_0$ is the frequency of the $F = 4 \leftrightarrow F' = 3$ undisturbed transition. The dots are the experiment, the solid lines are the approximation of the experimental data using Eq.~\eqref{eq:16}.}\label{fig6}
\end{figure*}

\begin{figure}[!t]  
	\includegraphics[width=\linewidth]{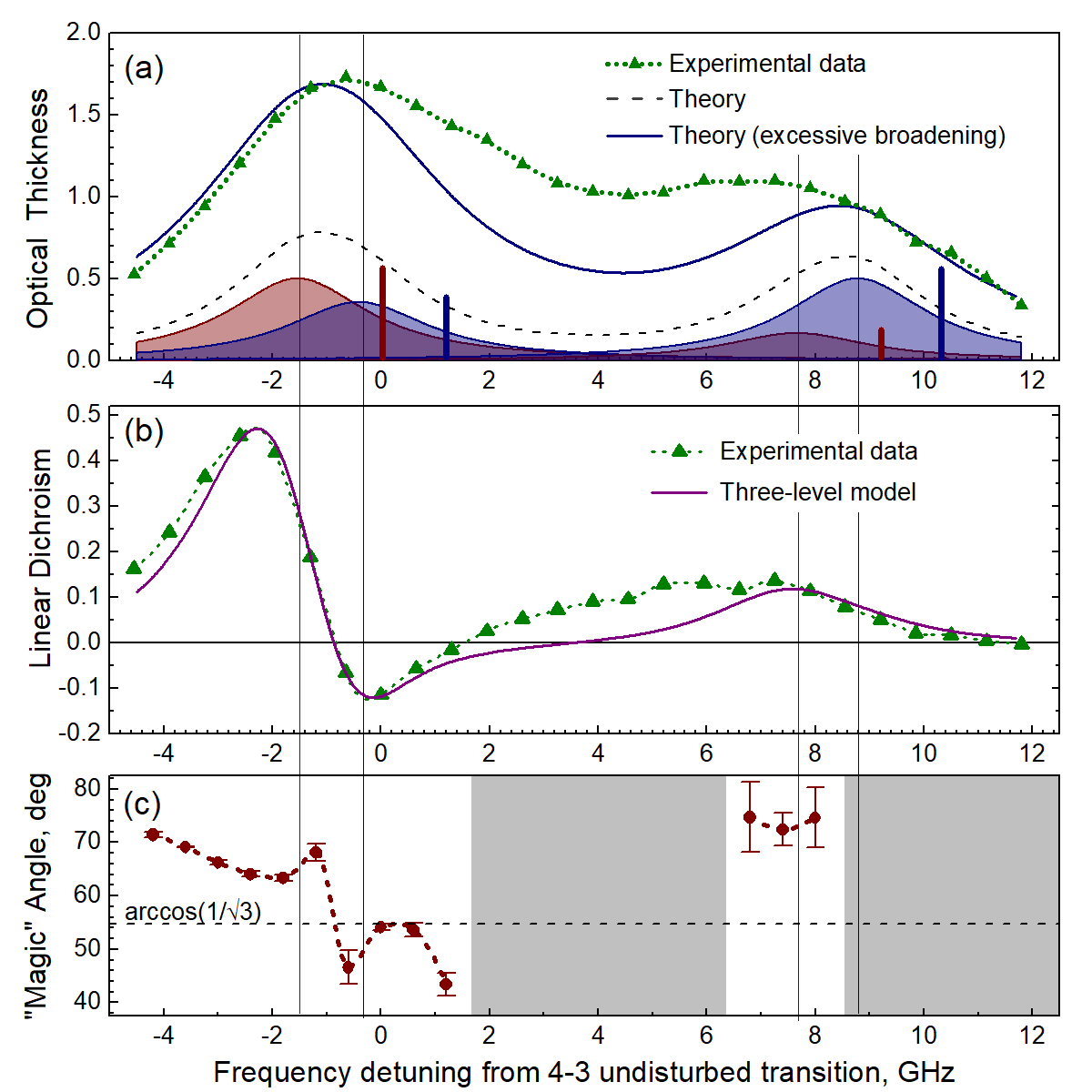}
	\caption{(a) The optical thickness normalized by the angle-averaged optical thickness. The dotted line is the calculated dependence of the Cs absorption spectrum (D\textsubscript{1} line) in a cell filled with 200~Torr N\textsubscript{2}. Individual optical profiles are filled with colors; triangles are the optical thickness measured at a light power of 8~mW; the solid line is the calculation taking into account the broadening due to the bleaching of the medium. Vertical lines in (a)-(c) indicate the calculated positions of the transition centers shifted by the buffer gas pressure. (b) Dependence of LD on the laser frequency: triangles are the experiment, the solid line is the calculation using model Eq.~\eqref{eq:33} at $(r_1 - r_2)^2 = 0.045$ for the left pair of transitions. (c) The ``magic'' angle (experimental). Gray rectangles indicate regions where the ``magic'' angle is absent. }\label{fig7}
\end{figure}

\begin{figure}[!t]  
	\includegraphics[width=\linewidth]{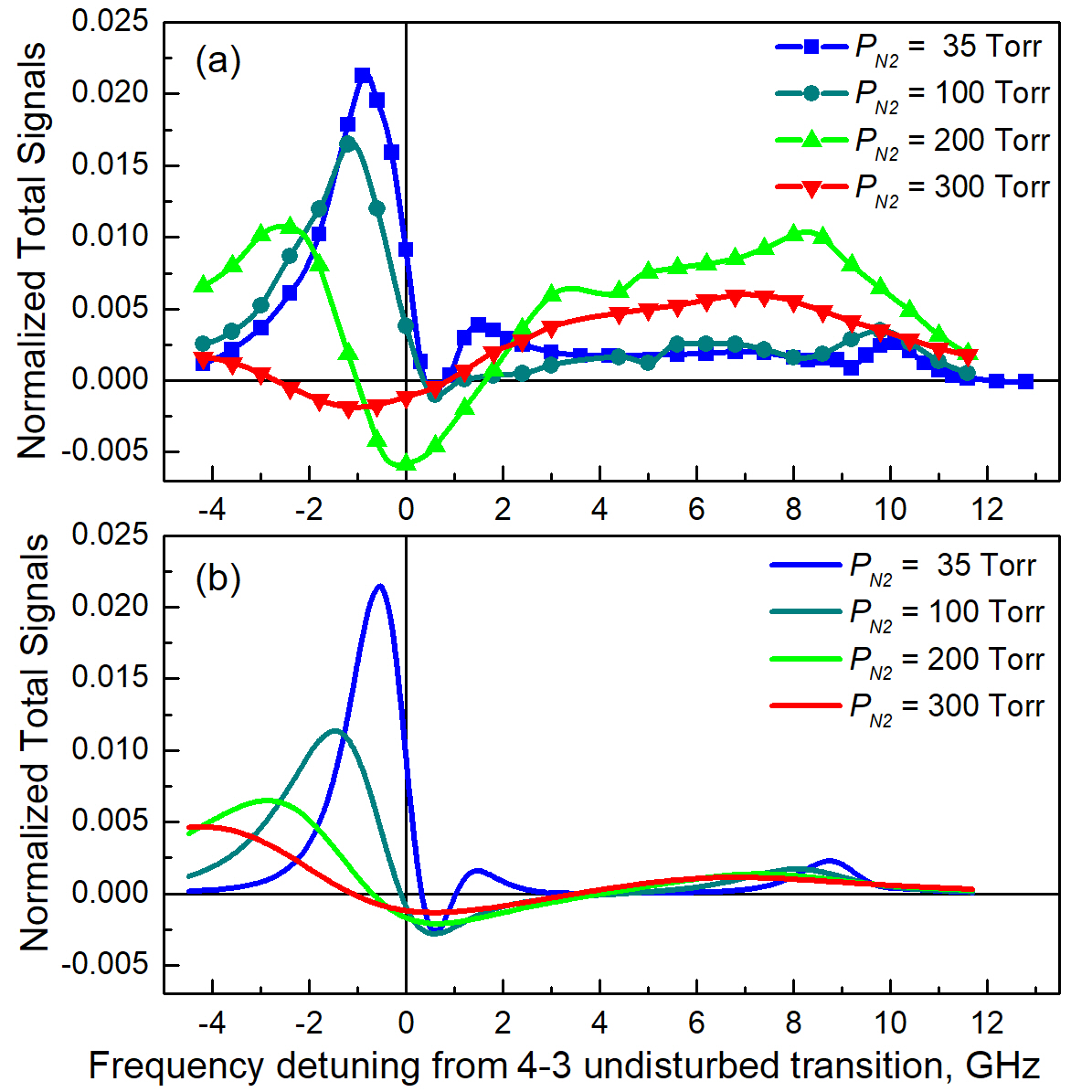}
	\caption{Dependence of LD on laser frequency in four cells with different buffer gas (nitrogen) pressures: (a) the experiment, (b) the calculation using model Eq.~\eqref{eq:33} at the same conditions as in Figure \ref{fig7}.}\label{fig8}
\end{figure}

Figures \ref{fig6} -- \ref{fig8} show selectively the experimental results and their comparison with the predictions of the model considered above. Figure \ref{fig6} shows four dependencies of the total and balanced signals on the azimuth of the input light polarization, selected from twenty-eight records taken at different frequency detunings of the laser light. The results of their approximation by Eq.~\eqref{eq:16} when selecting the values of $\Delta_x$ and $\Delta_y$ are also shown.

Figure \ref{fig7}a shows: the calculated dependence of the Cs absorption spectrum on the D1 line broadened by collisions with gas (gray dotted line, individual contours are filled in color); the experimentally measured optical thickness (triangles); the correction of the calculation taking into account the additional broadening arising from the non-uniform bleaching of the medium (solid line). Bleaching is a complex nonlinear process, which in alkali metals occurs mainly due to hyperfine pumping and leads to a nonlinear distortion of the absorption contours, which in the first order can be described by additional broadening. 

Figure \ref{fig7}b shows: the experimentally measured LD signal depending on the laser frequency (triangles) and the calculation according to model Eq.~\eqref{eq:33} (solid line). When calculating for the left pair of levels, the coefficients $\Delta_{\text{M}1} = -\Delta_{\text{M}2}$ and $r_1-r_2 = 0.21$ (corresponding to $(r_1-r_2)^2$ = 0.045) were taken.

In Figure \ref{fig7}b, the approximation for the left (low-frequency) pair of transitions, including the section with inverted dichroism, can be considered satisfactory. The task of exact approximation of the right pair of levels was not set at this stage, since, due to the greater statistical weight of the “dark'' level, they are characterized by a significantly lower value of dichroism (see Figure \ref{fig4}). 

Figure \ref{fig7}c illustrates the assumption that in systems with a complex spectrum the value of the “magic'' angle can deviate from its “classical'' value $\varphi_{\text{M}} = \arccos(1/\sqrt{3})$. It is evident that at some laser frequencies the differences of the “magic'' angle from its “classical'' value can significantly exceed the experimental error; moreover, in some frequency ranges the “magic'' angle is not observed at all. It can also be seen that the sharpest feature in the dependence, which has the form of a dispersion curve against a sloping baseline, is located in the region of maximum overlap of adjacent transitions $F = 4 \leftrightarrow F' = 3, 4$. Similar dependencies were observed in all studied cells.

Figure \ref{fig8} shows a comparison of the LD measurement results in four cells with different nitrogen pressures with the predictions of the same model. It is evident that for the left pair of transitions, good qualitative agreement between theory and experiment is maintained, including the dependence of dichroism zeroing points on laser frequency and cell pressure.

\section{DISCUSSION}\label{sec:5}

The results presented in the previous section indicate that the theory based on simple models can successfully describe such complex atoms as the cesium atom. In the previous section, a comparison of this theory with experiment was given, and satisfactory agreement was demonstrated. However, it should be noted that the dependences in demonstrate noticeable deviations of the angle at which the dichroism signal is zeroed from the ``magic''  angle. 

One explanation (as given above) is the unequal conservation of the nuclear spin in the process of pumping-decay of the excited state in terms of the magnetic quantum number. At the same time, Figure \ref{fig6} demonstrates very good agreement between the theory and experiment for the signals of both types (absorption and polarization rotation) when we apply the values of $\Delta_x$ and $\Delta_y$ obtained by fitting method. However, we have not yet performed a theoretical calculation of the observed values of $\Delta_x$ and $\Delta_y$ in a complex spectrum with two pairs of overlapping transitions -- this is a task for the next stage of the work, since it requires going beyond the simple models adopted in this article. Figure \ref{fig7}b also demonstrates the applicability of the theory to the description of cesium -- it compares the directly measured dichroism coefficient with the coefficient calculated by Eq.~\eqref{eq:16}, based on the values of $\Delta_x$ and $\Delta_y$ obtained during the approximation partially shown in Figure \ref{fig6}. Good agreement is also observed between both data sets and the calculation using model Eq.~\eqref{eq:33}, which confirms our assumptions about the inequality of the pump and detection efficiencies and the role of crossover transitions. It is also interesting to compare the predictions of the above theory with the results of a numerical calculation that takes into account all the features of the cesium structure. In Appendix~C we present matrices $\mathbf{a}$ and $\mathbf{b}$ for this model, obtained in the first order of expansion in intensity. By comparing them with the predictions of our simple models, we can conclude that not only is the approach proposed in this paper applicable to the model of a real multilevel atom, but so are the regularities already present in the model with three levels interacting with light. This is logical, since we are considering an alignment to which the populations of three groups of levels (the group of central levels and two groups of extreme levels) always contribute, and which is not affected by the features of the distribution of populations on a smaller scale (the latter further confirmed by the theory of the spherical representation).

\section{CONCLUSIONS}\label{sec:6}

Thus, we have related the internal properties of the medium with the angular dependences of absorption under optical pumping, separately considering the absorption in the single-beam scheme, which is of greatest practical interest. We showed that an anisotropic absorption in the single-beam scheme can be decomposed into components with a “dipole'' and “quadrupole'' (using the terminology of \cite{Fomin_Kozlov_Petrov_Smirnov_Petrenko_Zapasskii_2025}) angular dependence.

It is important to note that the “quadrupole'' angular dependence of absorption in the single-beam scheme is not a result of pumping nonlinearity: it is present even in the weak-light approximation. However, this is not a property of the medium (the absorption coefficient for weak probe light with a fixed polarization demonstrates a purely “dipole'' dependence), but a result of the anisotropic interaction of the aligned medium with the detection light. We also showed that in the general case the LD-determining properties of the medium can be characterized by only two parameters $\Delta_x$, $\Delta_y$, and in the case of complete mixing -- by a single parameter $\Delta_{\text{M}}$. We demonstrated the applicability of this theory, based on simple models, to more complex atomic systems, such as $^{133}$Cs, the ground state of which contains sixteen Zeeman sublevels.

We also showed experimentally that in systems with a complex spectrum, the value of the ``magic''  angle can vary. The sharpest changes correspond to the regions of overlap of adjacent optical transitions, which also suggests a connection between this phenomenon and ``cross''  detection.

The significant result of the work is the proof that incomplete mixing of the Zeeman sublevels of the excited state under optical pumping can lead to the inversion of dichroism signals, as well as expressions for calculating the magnitude of this phenomenon. The practical significance of this effect is that it is the basis for the principle of our recently proposed method of laser light stabilization 
\cite{Petrenko_Pazgalev_Vershovskii_2024}. This method, which, unlike existing methods of stabilization by atomic optical transitions, does not require either modulation of laser frequency or the use of strong magnetic fields, is potentially capable of replacing existing methods in problems of cooling atoms, atomic interferometry, optical frequency standards -- that is, wherever stable monochromatic laser light is required, and the use of powerful magnets is unacceptable.

\textbf{}

\textbf{Conflict of interest:} The authors declare that they have no conflict of interest.

\textbf{}

\textbf{Funding:} This research was funded by the baseline project FFUG-2024-0039 at the Ioffe Institute.

\textbf{}

\textbf{Acknowledgments:} The authors thank Prof.~Eugene B.~Aleksandrov and Prof.~Valery S. Zapasskii for valuable discussions.
\textbf{}

\section*{APPENDIX A: THE SYSTEM OF BALANCE EQUATIONS}
\renewcommand{\theequation}{A\arabic{equation}}
\setcounter{equation}{0}

\begin{table*}[!t]
\caption{\label{tab:t1}Optical pumping rate coefficients $P_{mF,mF'}$ for magnetic sublevels of the ground state of atoms with $F = 1$.}
\begin{ruledtabular}
\begin{tabular}{c|c|c|ccc|ccccc}
 & &{${F' = 0}$}& &{${F' = 1}$}& & & &{${F' = 2}$}& & \\ \hline
 &{$m_F$}& ${m_F}'=0$ &  ${m_F}'=-1$ & ${m_F}'=0$ &  ${m_F}'=1$ & ${m_F}'=-2$ & ${m_F}'=-1$ & ${m_F}'=0$ & ${m_F}'=1$ & ${m_F}'=2$ \\  \hline
 &   --1 & 1 & 1/2 & 1/2 & 0 & 1/10 & 3/10 & 6/10 & 0 & 0 \\ 
{$F = 1$} & 0 & 1 & 1/2 & 0 & 1/2 & 0 & 3/10 & 4/10 & 3/10 & 0 \\ 
 & 1 & 1 & 0 & 1/2 & 1/2 & 0 & 0 & 6/10 & 3/10 & 1/10 \\ 
\end{tabular}
\end{ruledtabular}
\end{table*}

In the general case, for an atomic system similar to the one shown in Figure \ref{fig1} (with a ground state with $N=2(2I+1)$ Zeeman levels), the system of balance equations looks like this:

\begin{equation}
\left\{\begin{array}{l} {\frac{\displaystyle dn_{F,m_{F} } }{\displaystyle dt} =Q-(R_{F,m_{F} } +{\rm \gamma })n_{F,m_{F} }} \\ {\cdots} \\ {\sum _{F=I-1/2}^{I+1/2}\sum _{m_{F} =-F}^{F}n_{F,m_{F} }   =1,} \end{array} \right. 
\label{eq:A1}
\end{equation}
where $n_{F,m_F}$ is the population of level $|F,mF>$, $R_{F,mF}$ is the total rate of depletion of level $|F,mF>$, $Q$ is the total rate of level population due to relaxation of both the excited and ground states:
\begin{equation}
\begin{array}{l} {R_{_{F,m_{F} } } =\gamma (\sum _{F'}^{}\sum _{m_{F} '}^{}P_{F,F',m_{F} ,m_{F'} } r_{F,F',m_{F} ,m_{F'} }   ),} \\ {} \\{Q=\frac{\displaystyle 1}{\displaystyle N} (\sum _{F}^{}\sum _{m_{F} }^{}R_{F,m_{F} } n_{F,m_{F} } +{\rm \gamma }  ),} 
\end{array}
\label{eq:A2}
\end{equation}
where $\gamma$ is the relaxation rate of the ground state levels; $P_{F,F',mF,mF'}$ are the relative probabilities of absorption of light of unit intensity and the corresponding polarization by the ground state sublevels; $r_{F,F',mF,mF'}$ are the dimensionless (normalized to $\gamma$) effective pump rates.

Transitions without a change in the magnetic quantum number ($\Delta m_F = 0$) occur under the influence of the light component linearly polarized along x ($r_{F,F',mF,mF} = I_x$), transitions with a change in the magnetic quantum number ($\Delta m_F = \pm1$) occur under the influence of the corresponding circular light component ($r_{F,F',mF,mF\pm1} = I\sigma_{\pm}$); for  $\Delta m_F > 1$ $P_{F,F',mF,mF'} = 0$. The method for calculating the values of $P_{F,F',mF,mF'}$ is given in 
\cite{Alexandrov_Chaika_Khvostenko_1993}. Tabular values of relative absorption probabilities for transitions in the structure of D\textsubscript{1} and $D_2$ lines of alkali metals are given in \cite{Varshalovich_Moskalev_Khersonskii_1988}.

Solution of Eq.~\eqref{eq:A1} can be sought for the stationary case; for this the time derivatives in Eq.~\eqref{eq:A1} are equated to zero. Further calculation of absorption can be performed in two ways: either directly, by summing up the absorption (using the same probability coefficients) from all levels involved, or by decomposing the polarization of the medium into spherical moments with subsequent multiplication of the polarization moments of the density matrix of the medium by the corresponding moments of the polarization matrix of light. Multipole moments are a special case of decomposition of the density matrix of the medium into irreducible tensor components 
\cite{Blum_2012,Alem_Hughes_Buard_Cheung_Maydew_Griesshammer_Holloway_Park_Lechuga_Coolidge_2023}. The application of these two methods should lead to the same results, but the second method is convenient in that it allows us to separate the alignment (second-order moment) from the orientation (first-order moment) and higher-order moments.

All three models shown in Figure \ref{fig2} can be described by the following system of balance equations:
\begin{equation}
\left\{\begin{array}{l} {\frac{\displaystyle dn_{1} }{\displaystyle dt} =Q-(R_{1} +{\rm \gamma })n_{1} } \\{}\\  {\frac{\displaystyle dn_{2} }{\displaystyle dt} =Q-(R_{2} +{\rm \gamma })n_{2} } \\{}\\  {\frac{\displaystyle dn_{d} }{\displaystyle dt} =Q-{\rm \gamma }n_{d} } \\{}\\  {gn_{d} +n_{1} +2n_{2} =1,} \end{array} \right .
\label{eq:A3}
\end{equation}
where 
\begin{equation}
\begin{array}{l} {R_{1} =P_{1,1} I_{x} +P_{1,2} I_{{\rm \sigma }-} +P_{1,3} I_{{\rm \sigma }+} ,} \\{}\\ {R_{2} =P_{2,2} I_{x} +P_{2,4} I_{{\rm \sigma }-} +P_{2,1} I_{{\rm \sigma }+} ,} \\{}\\ {Q=\frac{1}{3} (R_{1} n_{1} +2R_{2} n_{2} +\gamma ).} 
\end{array}
\label{eq:A4}
\end{equation}

In what follows, we will use dimensionless pump intensities normalized to $\gamma$. According to 
\cite{Budker_Kimball_DeMille_2004} (p.375), the alignment in the stationary case can be described by a single quantity -- the multipole $\rho^{(2)}_{0} \sim <3F^2_z - F^2>$, in a three-level system equal to
\begin{equation}
{\rm \rho }_{{\rm 0}}^{{(2)}} =\frac{1}{\sqrt{6} } (n_{2} -2n_{1} +n_{3} )=\frac{1}{\sqrt{6} } (2n_{2} -2n_{1} ).
\label{eq:A5}
\end{equation}

In real atoms, at the dark hyperfine level, as a result of spin exchange processes, alignment can also be created as a result of the transfer of alignment, but since the ``dark''  level does not interact with light, it does not manifest itself directly in absorption. The calculation of absorption is reduced to the expressions
\begin{equation}
\begin{array}{l} {K_{x} =K_{0} (P_{1,1} n_{1} +2P_{2,2} n_{2} )}, \\ {} \\
{K_{y} =\frac{\displaystyle 1}{\displaystyle 2} K_{0} (2P_{1,2} n_{1} +(P_{2,1} +P_{2,4} )n_{2}} \\{}\\
\qquad +(P_{3,1} +P_{3,5}) n_{3} ). 
\end{array}
\label{eq:A6}
\end{equation}

Note that this definition implies that the probe light is characterized by the same spectral composition as the pumping light. The cumbersome solutions of system Eq.~\eqref{eq:A3} can be significantly simplified if we move from the probabilities of specific transitions to the probabilities of absorption of the pump light components:
\begin{equation}
\begin{array}{l} {\mathbf{p}=\left(\begin{array}{cc} {p_{c,x} } & {p_{c,y} } \\ {p_{e,x} } & {p_{e,y} } \end{array}\right)=\left(\begin{array}{cc} {P_{1,1} } & {P_{1,2} } \\ {P_{2,2} } & {\frac{\displaystyle 1}{\displaystyle 2} (P_{2,1} +P_{2,4} )} \end{array}\right)} \\{}\\ {\qquad \equiv \left(\begin{array}{cc} {p_{x0} +2\Delta _{x} } & {p_{y0} +2\Delta _{y} } \\ {p_{x0} -\Delta _{x} } & {p_{y0} -\Delta _{y} } \end{array}\right),} 
\end{array}
\label{eq:A7}
\end{equation}

\begin{equation}
\begin{array}{cc} {\Delta _{x} =\frac{\displaystyle P_{1,1} -P_{2,2} }{\displaystyle 3} ;} & {\Delta _{y} =\frac{\displaystyle 2P_{1,2} -P_{2,1} -P_{2,4} }{\displaystyle 6} ;} \\ {} \\ {p_{x0} =\frac{\displaystyle P_{1,1} +2P_{2,2} }{\displaystyle 3} ;} & {p_{y0} =\frac{\displaystyle 2P_{1,2} +P_{2,1} +P_{2,4} }{\displaystyle 3} .}
\end{array}
\label{eq:A8}
\end{equation}

The indices $c$, $e$ correspond to the central and extreme levels, which give, respectively, a negative and positive contribution to Eq.~\eqref{eq:A5}; for a three-level scheme $c=1$ ($m_F~=~0$), and $e=2, 3$ $(m_F=\pm1)$. The coefficients $p_i,j$ form a 2$\times$2 matrix $\mathbf{p}$. For the transitions in the schemes presented in Figure \ref{fig2}, these coefficients are given in Table~\ref{tab:t2}.

For the case of real atomic systems, expressions Eq.~\eqref{eq:A7}, \eqref{eq:7} can be simplified. First, we see that the ratio $p_{x0}/p_{y0}$ determines dichroism in the absence of pumping; therefore, in the case of the induced dichroism $p_{x0}=p_{y0}$. Since the pumping intensities are normalized to the lifetime of one magnetic sublevel $\gamma=1$, the average rate of optical pumping from the sublevels interacting with light at a unit total light intensity $I_0 = I_y+I_x$ must be equal to unity:
\begin{equation}
\frac{1}{3} (p_{1,x} I_{x} +2p_{2,x} I_{x} +p_{1,y} I_{y} +2p_{2,y} I_{y} )=1. 
\label{eq:A9}
\end{equation}

 Substituting into Eq.~\eqref{eq:9} $I_y = 2I_x$ (which corresponds to isotropic pumping with unpolarized light) and $I_y+I_x=1$, and taking into account Eqs.~\eqref{eq:A7}, \eqref{eq:7}, we obtain $p_{x0}=p_{y0}=1$.

\section*{APPENDIX B: THE TOTAL AND BALANCED SIGNALS}
\renewcommand{\theequation}{B\arabic{equation}}
\setcounter{equation}{0}

We will start with signals expressed in absolute intensity units and move on to normalized values. The total $S_T$ signal measured in transmitted light is simply equal to the intensity at the cell output. The magnitude of the $S_B$ signal, equal to the difference in intensities in two perpendicular polarizations, is determined by the angle $\Delta \varphi = \varphi - \varphi_0$ of rotation of the beam polarization plane signal, 
\begin{equation}
S_{B} =-\frac{I}{2} \sin (2\Delta \varphi )\approx -I\Delta \varphi .
\label{eq:B1}
\end{equation}

At $\Delta \varphi \ll1$, which is always the case in the case of LD signals, $S_B \sim \Delta \varphi$. The theoretical value of the $S_B$ signal can be calculated by substituting the coefficients Eq.~\eqref{eq:13} into Eqs.~\eqref{eq:1}, \eqref{eq:2} followed by rotating the coordinate system by $\pi/4 - \varphi$ and calculating the difference in the intensities of the corresponding polarization components:
\begin{equation} 
\left(\begin{array}{c} {E_{Rx'' } } \\ {E_{Ry'' } } \end{array}\right)
=\mathbf{R}(\frac{\pi }{4} -\varphi )\cdot \mathbf{T}\cdot \left(\begin{array}{c} {E_{x} } \\ {E_{y} } \end{array}\right),
\label{eq:B2}
\end{equation}
where $\mathbf{R}(\psi $) is the matrix of rotation by the angle $\psi $. Then
\begin{equation}
S_{B} =E_{Ry'' }^{*} E_{Ry'' } -E_{Rx'' }^{*} E_{Rx'' }.
\label{eq:B3}
\end{equation}

Thus, for normalized signals $S_{TN}=S_T/I_0$ and $S_{BN}=S_B/I_0$ we obtain
\begin{equation}
\begin{array}{l} {S_{TN} =1-K=(1-\bar{K})+\Delta K\cos (2\varphi ),} \\ {} \\ {S_{BN} =\Delta K\cdot \sin (2\varphi )} \\{}\\ {\qquad +\frac{\displaystyle 1-\bar{K}}{\displaystyle 2} (1-\sqrt{1-(\frac{\displaystyle \Delta K}{\displaystyle 1-\bar{K}} )^{2} } )\sin (4\varphi ).} 
\label{eq:B4}
\end{array}
\end{equation}

We can also express these signals explicitly through Eq.~\eqref{eq:12}, and extract the full angular dependences in them. Neglecting the second term in the expression for $S_B$ (thin layer approximation $\Delta K\ll 1-K)$, we obtain,
\begin{equation}
\begin{array}{l} {S_{TN} =1-\frac{K_{0} }{1+{\rm \beta }I_{0} } (1+\frac{1}{q} I_{0} w),} \\{}\\ {S_{BN} =2\Delta_{-} \frac{K_{0} }{1+{\rm \beta }I_{0} } I_{0} v\sin (2\varphi ).} 
\end{array}
\label{eq:B5}
\end{equation}

\begin{table*}[!t]
\centering
\caption{\label{tab:t2}Coefficients $p_i$,$j$, $\Delta_x$, $\Delta_y$, $\Delta_-$, $\Delta_+$ for magnetic sublevels of the ground state of atoms with $F = 1$.}
\begin{tabular}{c| c |cc|cc|cc}
\hline
\hline
 & &\multicolumn{2}{c|}{${F' = 0}$}&\multicolumn{2}{c|}{${F' = 1}$}&\multicolumn{2}{c}{${F' = 2}$} \\ \hline
 & { $\quad i\quad $ }& $\quad j=x\quad $ & $\quad j=y\quad $ &  $\quad j=x\quad $ &  $\quad j=y\quad $ &  $\quad j=x\quad $ &  $\quad j=y\quad $ \\ \hline
 & 1 & 1 & 0 & 0 & 1/2 & 4/10 & 3/10 \\ 
 $\quad p_{i,j},\quad$ & 2 & 0 & 1/2 & 1/2 & 1/4 & 3/10 & 7/20 \\ \cline{2-8}
 $\quad F=1\quad $& 1 & 3 & 0 & 0 & 3/2 & 12/10 & 9/10 \\ 
 & 2 & 0 & 3/2 & 3/2 & 3/4 & 9/10 & 21/20 \\ \hline
 $\Delta $x &  & \multicolumn{2}{c|}{1} & \multicolumn{2}{c|}{--1/2} & \multicolumn{2}{c}{1/10} \\ 
$\Delta $y &  & \multicolumn{2}{c|}{--1/2} & \multicolumn{2}{c|}{1/4} & \multicolumn{2}{c}{--1/20} \\ 
$\Delta $- &  & \multicolumn{2}{c|}{3/4} & \multicolumn{2}{c|}{--3/8} & \multicolumn{2}{c}{3/40} \\ 
$\Delta $+ &  & \multicolumn{2}{c|}{1/4} & \multicolumn{2}{c|}{--1/8} & \multicolumn{2}{c}{1/40} \\ 
\hline
\hline
\end{tabular}
\end{table*}

\section*{APPENDIX C: THE “MAGIC'' ANGLE}
\renewcommand{\theequation}{C\arabic{equation}}
\setcounter{equation}{0}
Taking into account the “magic'' angle condition Eq.~\eqref{eq:21}, Eqs.~\eqref{eq:9}, \eqref{eq:11} are reduced to
\begin{equation}
\begin{array}{l} {\mathbf{a}=q\left(\begin{array}{cc} {1} & {1} \\ {1} & {1} \end{array}\right)+\Delta _{{\rm M}}^{{2}} \left(\begin{array}{cc} {2} & {1} \\ {1} & {1/2} \end{array}\right),} \\{}\\ {\mathbf{b}=(1+q)\left(\begin{array}{c} {1} \\ {1} \end{array}\right)+\Delta _{{\rm M}} \left(\begin{array}{c} {1} \\ {-1/2} \end{array}\right),} \\ {} \\ {c_{0} =2qI^{2} +\Delta _{{\rm M}} (I_{x}^{2} +I_{x} I_{y} -I_{y}^{2} )} \\{}\\ {\qquad +q\Delta _{{\rm M}}^{{\rm 2}} (-4I_{x}^{2} +4I_{x} I_{y} -I_{y}^{2} ).} \end{array}
\label{eq:C1}
\end{equation}

Eq.~\eqref{eq:14} now takes the form
\begin{equation} 
{\rho }_{0}^{(2)} =\sqrt{\frac{2}{3} } I_{0} \frac{(1-q)v_{0} }{1+I_{0} (1+q+v_{0} )+I_{0}^{2} w_{0} } ,
\label{eq:C2}
\end{equation}
where 
\begin{equation}
\begin{array}{l} {v_{0} =\frac{\displaystyle \Delta _{{\rm M}} }{\displaystyle 4} (1+3\cos (2\varphi ))=\Delta _{{\rm M}} P_{L2} ,} \\ {} \\ {w_{0} =q(1-v_{0} )(1+2v_{0} ).} \end{array}
\label{eq:C3}
\end{equation}

We also present here a comparison with the result of a full-fledged numerical calculation, taking into account all 16 levels of the ground state and 16 levels of the excited state of cesium (D\textsubscript{1} line) under the influence of linearly polarized pumping. The calculated values of the coefficients $K_x$ and $K_y$ were reduced to the form Eq.\eqref{eq:8}, and as a result of comparison with Eq.\eqref{eq:8}, the matrices $\mathbf{a}_{\text{Cs}}$ and $\mathbf{b}_{\text{Cs}}$ were calculated:
\begin{equation}
\begin{array}{cccc} {\mathbf{a}_{\tt{Cs}} =\left(\begin{array}{cc} {\frac{31}{21} } & {\frac{3}{7} } \\ {\frac{3}{7} } & {\frac{20}{21} } \end{array}\right)} & {\mathbf{b}_{\tt{Cs}} =\left(\begin{array}{cc} {\frac{179}{21} } & {\frac{53}{7} } \\ {\frac{179}{21} } & {\frac{53}{7} } \end{array}\right).} & {} & {} 
\end{array}
\label{eq:C4}
\end{equation}
Note that the components of the $\mathbf{a}_{\text{Cs}}$ matrix exactly correspond to Eq.~\eqref{eq:C1} if we set $\Delta = \sqrt{(22/63)}, g =112/9$. The components of the $\mathbf{b}_{\text{Cs}}$ matrix correspond to Eq.~\eqref{eq:C1} if we set $\Delta = 40/63, g = 1109/18$.




%


\begin{thebibliography}{33}%
\makeatletter
\providecommand \@ifxundefined [1]{%
 \@ifx{#1\undefined}
}%
\providecommand \@ifnum [1]{%
 \ifnum #1\expandafter \@firstoftwo
 \else \expandafter \@secondoftwo
 \fi
}%
\providecommand \@ifx [1]{%
 \ifx #1\expandafter \@firstoftwo
 \else \expandafter \@secondoftwo
 \fi
}%
\providecommand \natexlab [1]{#1}%
\providecommand \enquote  [1]{``#1''}%
\providecommand \bibnamefont  [1]{#1}%
\providecommand \bibfnamefont [1]{#1}%
\providecommand \citenamefont [1]{#1}%
\providecommand \href@noop [0]{\@secondoftwo}%

\providecommand \@href[1]{\@@startlink{#1}\@@href}%
\providecommand \@@href[1]{\endgroup#1\@@endlink}%
\providecommand \@sanitize@url [0]{\catcode `\\12\catcode `\$12\catcode `\&12\catcode `\#12\catcode `\^12\catcode `\_12\catcode `\%12\relax}%
\providecommand \@@startlink[1]{}%
\providecommand \@@endlink[0]{}%

\providecommand \@url [1]{\endgroup\@href {#1}{\urlprefix }}%
\providecommand \urlprefix  [0]{URL }%

\providecommand \doibase [0]{https://doi.org/}%
\providecommand \selectlanguage [0]{\@gobble}%
\providecommand \bibinfo  [0]{\@secondoftwo}%
\providecommand \bibfield  [0]{\@secondoftwo}%
\providecommand \translation [1]{[#1]}%
\providecommand \BibitemOpen [0]{}%
\providecommand \bibitemStop [0]{}%
\providecommand \bibitemNoStop [0]{.\EOS\space}%
\providecommand \EOS [0]{\spacefactor3000\relax}%
\providecommand \BibitemShut  [1]{\csname bibitem#1\endcsname}%
\let\auto@bib@innerbib\@empty
\bibitem [{\citenamefont {Budker}\ and\ \citenamefont {Romalis}(2007)}]{Budker_Romalis_2007}%
  \BibitemOpen
  \bibfield  {author} {\bibinfo {author} {\bibfnamefont {D.}~\bibnamefont {Budker}}\ and\ \bibinfo {author} {\bibfnamefont {M.}~\bibnamefont {Romalis}},\ }\bibfield  {title} {\bibinfo {title} {Optical magnetometry},\ }\href {https://doi.org/10.1038/nphys566} {\bibfield  {journal} {\bibinfo  {journal} {Nature Physics}\ }\textbf {\bibinfo {volume} {3}},\ \bibinfo {pages} {227} (\bibinfo {year} {2007})}\BibitemShut {NoStop}%
\bibitem [{\citenamefont {Romalis}(2022)}]{Romalis_2022}%
  \BibitemOpen
  \bibfield  {author} {\bibinfo {author} {\bibfnamefont {M.~V.}\ \bibnamefont {Romalis}},\ }\bibinfo {title} {Optically pumped magnetometers for biomagnetic measurements},\ in\ \href {https://doi.org/10.1007/978-3-031-05363-4_1} {\emph {\bibinfo {booktitle} {Flexible High Performance Magnetic Field Sensors: On-Scalp Magnetoencephalography and Other Applications}}},\ \bibinfo {editor} {edited by\ \bibinfo {editor} {\bibfnamefont {E.}~\bibnamefont {Labyt}}, \bibinfo {editor} {\bibfnamefont {T.}~\bibnamefont {Sander}},\ and\ \bibinfo {editor} {\bibfnamefont {R.}~\bibnamefont {Wakai}}}\ (\bibinfo  {publisher} {Springer International Publishing},\ \bibinfo {address} {Cham},\ \bibinfo {year} {2022})\ p.\ \bibinfo {pages} {3–15}\BibitemShut {NoStop}%
\bibitem [{\citenamefont {Boto}\ \emph {et~al.}(2018)\citenamefont {Boto}, \citenamefont {Holmes}, \citenamefont {Leggett}, \citenamefont {Roberts}, \citenamefont {Shah}, \citenamefont {Meyer}, \citenamefont {Munoz}, \citenamefont {Mullinger}, \citenamefont {Tierney}, \citenamefont {Bestmann}, \citenamefont {Barnes}, \citenamefont {Bowtell},\ and\ \citenamefont {Brookes}}]{Boto_Holmes_Leggett_Roberts_Shah_Meyer_Munoz_Mullinger_Tierney_Bestmann_2018}%
  \BibitemOpen
  \bibfield  {author} {\bibinfo {author} {\bibfnamefont {E.}~\bibnamefont {Boto}}, \bibinfo {author} {\bibfnamefont {N.}~\bibnamefont {Holmes}}, \bibinfo {author} {\bibfnamefont {J.}~\bibnamefont {Leggett}}, \bibinfo {author} {\bibfnamefont {G.}~\bibnamefont {Roberts}}, \bibinfo {author} {\bibfnamefont {V.}~\bibnamefont {Shah}}, \bibinfo {author} {\bibfnamefont {S.~S.}\ \bibnamefont {Meyer}}, \bibinfo {author} {\bibfnamefont {L.~D.}\ \bibnamefont {Munoz}}, \bibinfo {author} {\bibfnamefont {K.~J.}\ \bibnamefont {Mullinger}}, \bibinfo {author} {\bibfnamefont {T.~M.}\ \bibnamefont {Tierney}}, \bibinfo {author} {\bibfnamefont {S.}~\bibnamefont {Bestmann}}, \bibinfo {author} {\bibfnamefont {G.~R.}\ \bibnamefont {Barnes}}, \bibinfo {author} {\bibfnamefont {R.}~\bibnamefont {Bowtell}},\ and\ \bibinfo {author} {\bibfnamefont {M.~J.}\ \bibnamefont {Brookes}},\ }\bibfield  {title} {\bibinfo {title} {Moving magnetoencephalography towards real-world applications with a wearable system},\ }\href
  {https://doi.org/10.1038/nature26147} {\bibfield  {journal} {\bibinfo  {journal} {Nature}\ }\textbf {\bibinfo {volume} {555}},\ \bibinfo {pages} {657} (\bibinfo {year} {2018})}\BibitemShut {NoStop}%
\bibitem [{\citenamefont {Knappe}\ \emph {et~al.}(2005)\citenamefont {Knappe}, \citenamefont {Schwindt}, \citenamefont {Shah}, \citenamefont {Hollberg}, \citenamefont {Kitching}, \citenamefont {Liew},\ and\ \citenamefont {Moreland}}]{Knappe_Schwindt_Shah_Hollberg_Kitching_Liew_Moreland_2005}%
  \BibitemOpen
  \bibfield  {author} {\bibinfo {author} {\bibfnamefont {S.}~\bibnamefont {Knappe}}, \bibinfo {author} {\bibfnamefont {P.}~\bibnamefont {Schwindt}}, \bibinfo {author} {\bibfnamefont {V.}~\bibnamefont {Shah}}, \bibinfo {author} {\bibfnamefont {L.}~\bibnamefont {Hollberg}}, \bibinfo {author} {\bibfnamefont {J.}~\bibnamefont {Kitching}}, \bibinfo {author} {\bibfnamefont {L.}~\bibnamefont {Liew}},\ and\ \bibinfo {author} {\bibfnamefont {J.}~\bibnamefont {Moreland}},\ }\bibfield  {title} {\bibinfo {title} {A chip-scale atomic clock based on 87 rb with improved frequency stability},\ }\href {https://doi.org/10.1364/OPEX.13.001249} {\bibfield  {journal} {\bibinfo  {journal} {Optics express}\ }\textbf {\bibinfo {volume} {13}},\ \bibinfo {pages} {1249–1253} (\bibinfo {year} {2005})}\BibitemShut {NoStop}%
\bibitem [{\citenamefont {Alem}\ \emph {et~al.}(2023)\citenamefont {Alem}, \citenamefont {Hughes}, \citenamefont {Buard}, \citenamefont {Cheung}, \citenamefont {Maydew}, \citenamefont {Griesshammer}, \citenamefont {Holloway}, \citenamefont {Park}, \citenamefont {Lechuga}, \citenamefont {Coolidge} \emph {et~al.}}]{Alem_Hughes_Buard_Cheung_Maydew_Griesshammer_Holloway_Park_Lechuga_Coolidge_2023}%
  \BibitemOpen
  \bibfield  {author} {\bibinfo {author} {\bibfnamefont {O.}~\bibnamefont {Alem}}, \bibinfo {author} {\bibfnamefont {K.~J.}\ \bibnamefont {Hughes}}, \bibinfo {author} {\bibfnamefont {I.}~\bibnamefont {Buard}}, \bibinfo {author} {\bibfnamefont {T.~P.}\ \bibnamefont {Cheung}}, \bibinfo {author} {\bibfnamefont {T.}~\bibnamefont {Maydew}}, \bibinfo {author} {\bibfnamefont {A.}~\bibnamefont {Griesshammer}}, \bibinfo {author} {\bibfnamefont {K.}~\bibnamefont {Holloway}}, \bibinfo {author} {\bibfnamefont {A.}~\bibnamefont {Park}}, \bibinfo {author} {\bibfnamefont {V.}~\bibnamefont {Lechuga}}, \bibinfo {author} {\bibfnamefont {C.}~\bibnamefont {Coolidge}}, \emph {et~al.},\ }\bibfield  {title} {\bibinfo {title} {An integrated full-head opm-meg system based on 128 zero-field sensors},\ }\href {https://doi.org/10.3389/fnins.2023.1190310} {\bibfield  {journal} {\bibinfo  {journal} {Frontiers in Neuroscience}\ }\textbf {\bibinfo {volume} {17}},\ \bibinfo {pages} {1014} (\bibinfo {year} {2023})}\BibitemShut {NoStop}%
\bibitem [{\citenamefont {Ludlow}\ \emph {et~al.}(2015)\citenamefont {Ludlow}, \citenamefont {Boyd}, \citenamefont {Ye}, \citenamefont {Peik},\ and\ \citenamefont {Schmidt}}]{Ludlow_Boyd_Ye_Peik_Schmidt_2015}%
  \BibitemOpen
  \bibfield  {author} {\bibinfo {author} {\bibfnamefont {A.~D.}\ \bibnamefont {Ludlow}}, \bibinfo {author} {\bibfnamefont {M.~M.}\ \bibnamefont {Boyd}}, \bibinfo {author} {\bibfnamefont {J.}~\bibnamefont {Ye}}, \bibinfo {author} {\bibfnamefont {E.}~\bibnamefont {Peik}},\ and\ \bibinfo {author} {\bibfnamefont {P.~O.}\ \bibnamefont {Schmidt}},\ }\bibfield  {title} {\bibinfo {title} {Optical atomic clocks},\ }\  {\bibfield  {journal} {\bibinfo  {journal} {Reviews of Modern Physics}\ }\textbf {\bibinfo {volume} {87}},\ \bibinfo {pages} {637–701} (\bibinfo {year} {2015})}\BibitemShut {NoStop}%
\bibitem [{\citenamefont {Meyer}\ and\ \citenamefont {Larsen}(2014)}]{Meyer_Larsen_2014}%
  \BibitemOpen
  \bibfield  {author} {\bibinfo {author} {\bibfnamefont {D.}~\bibnamefont {Meyer}}\ and\ \bibinfo {author} {\bibfnamefont {M.}~\bibnamefont {Larsen}},\ }\bibfield  {title} {\bibinfo {title} {Nuclear magnetic resonance gyro for inertial navigation},\ }\href {https://doi.org/10.1134/S2075108714020060} {\bibfield  {journal} {\bibinfo  {journal} {Gyroscopy and Navigation}\ }\textbf {\bibinfo {volume} {5}},\ \bibinfo {pages} {75–82} (\bibinfo {year} {2014})}\BibitemShut {NoStop}%
\bibitem [{\citenamefont {Sorensen}\ \emph {et~al.}(2020)\citenamefont {Sorensen}, \citenamefont {Thrasher},\ and\ \citenamefont {Walker}}]{Sorensen_Thrasher_Walker_2020}%
  \BibitemOpen
  \bibfield  {author} {\bibinfo {author} {\bibfnamefont {S.~S.}\ \bibnamefont {Sorensen}}, \bibinfo {author} {\bibfnamefont {D.~A.}\ \bibnamefont {Thrasher}},\ and\ \bibinfo {author} {\bibfnamefont {T.~G.}\ \bibnamefont {Walker}},\ }\bibfield  {title} {\bibinfo {title} {A synchronous spin-exchange optically pumped nmr-gyroscope},\ }\href {https://doi.org/10.3390/app10207099} {\bibfield  {journal} {\bibinfo  {journal} {Applied Sciences}\ }\textbf {\bibinfo {volume} {10}},\ \bibinfo {pages} {7099} (\bibinfo {year} {2020})}\BibitemShut {NoStop}%
\bibitem [{\citenamefont {Tino}(2021)}]{Tino_2021}%
  \BibitemOpen
  \bibfield  {author} {\bibinfo {author} {\bibfnamefont {G.~M.}\ \bibnamefont {Tino}},\ }\bibfield  {title} {\bibinfo {title} {Testing gravity with cold atom interferometry: results and prospects},\ }\href {https://doi.org/10.1088/2058-9565/abd83e} {\bibfield  {journal} {\bibinfo  {journal} {Quantum Science and Technology}\ }\textbf {\bibinfo {volume} {6}},\ \bibinfo {pages} {024014} (\bibinfo {year} {2021})}\BibitemShut {NoStop}%
\bibitem [{\citenamefont {Tino}\ and\ \citenamefont {Kasevich}(2014)}]{Tino_Kasevich_2014}%
  \BibitemOpen
  \bibfield  {author} {\bibinfo {author} {\bibfnamefont {G.~M.}\ \bibnamefont {Tino}}\ and\ \bibinfo {author} {\bibfnamefont {M.~A.}\ \bibnamefont {Kasevich}},\ } {\emph {\bibinfo {title} {Atom interferometry}}},\ Vol.\ \bibinfo {volume} {188}\ (\bibinfo  {publisher} {IOS Press},\ \bibinfo {year} {2014})\BibitemShut {NoStop}%
\bibitem [{\citenamefont {Paraiso}\ \emph {et~al.}(2021)\citenamefont {Paraiso}, \citenamefont {Woodward}, \citenamefont {Marangon}, \citenamefont {Lovic}, \citenamefont {Yuan},\ and\ \citenamefont {Shields}}]{Paraiso_Woodward_Marangon_Lovic_Yuan_Shields_2021}%
  \BibitemOpen
  \bibfield  {author} {\bibinfo {author} {\bibfnamefont {T.~K.}\ \bibnamefont {Paraiso}}, \bibinfo {author} {\bibfnamefont {R.~I.}\ \bibnamefont {Woodward}}, \bibinfo {author} {\bibfnamefont {D.~G.}\ \bibnamefont {Marangon}}, \bibinfo {author} {\bibfnamefont {V.}~\bibnamefont {Lovic}}, \bibinfo {author} {\bibfnamefont {Z.}~\bibnamefont {Yuan}},\ and\ \bibinfo {author} {\bibfnamefont {A.~J.}\ \bibnamefont {Shields}},\ }\bibfield  {title} {\bibinfo {title} {Advanced laser technology for quantum communications (tutorial review)},\ }\href {https://doi.org/10.1002/qute.202100062} {\bibfield  {journal} {\bibinfo  {journal} {Advanced Quantum Technologies}\ }\textbf {\bibinfo {volume} {4}},\ \bibinfo {pages} {2100062} (\bibinfo {year} {2021})}\BibitemShut {NoStop}%
\bibitem [{\citenamefont {Schreck}\ and\ \citenamefont {Druten}(2021)}]{Schreck_Druten_2021}%
  \BibitemOpen
  \bibfield  {author} {\bibinfo {author} {\bibfnamefont {F.}~\bibnamefont {Schreck}}\ and\ \bibinfo {author} {\bibfnamefont {K.~v.}\ \bibnamefont {Druten}},\ }\bibfield  {title} {\bibinfo {title} {Laser cooling for quantum gases},\ }\href {https://doi.org/10.1038/s41567-021-01379-w} {\bibfield  {journal} {\bibinfo  {journal} {Nature Physics}\ }\textbf {\bibinfo {volume} {17}},\ \bibinfo {pages} {1296–1304} (\bibinfo {year} {2021})}\BibitemShut {NoStop}%
\bibitem [{\citenamefont {Nagel}\ \emph {et~al.}(1998)\citenamefont {Nagel}, \citenamefont {Graf}, \citenamefont {Naumov}, \citenamefont {Mariotti}, \citenamefont {Biancalana}, \citenamefont {Meschede},\ and\ \citenamefont {Wynands}}]{Nagel_Graf_Naumov_Mariotti_Biancalana_Meschede_Wynands_1998}%
  \BibitemOpen
  \bibfield  {author} {\bibinfo {author} {\bibfnamefont {A.}~\bibnamefont {Nagel}}, \bibinfo {author} {\bibfnamefont {L.}~\bibnamefont {Graf}}, \bibinfo {author} {\bibfnamefont {A.}~\bibnamefont {Naumov}}, \bibinfo {author} {\bibfnamefont {E.}~\bibnamefont {Mariotti}}, \bibinfo {author} {\bibfnamefont {V.}~\bibnamefont {Biancalana}}, \bibinfo {author} {\bibfnamefont {D.}~\bibnamefont {Meschede}},\ and\ \bibinfo {author} {\bibfnamefont {R.}~\bibnamefont {Wynands}},\ }\bibfield  {title} {\bibinfo {title} {Experimental realization of coherent dark-state magnetometers},\ }\href {https://doi.org/10.1209/epl/i1998-00430-0} {\bibfield  {journal} {\bibinfo  {journal} {Europhysics Letters}\ }\textbf {\bibinfo {volume} {44}},\ \bibinfo {pages} {31} (\bibinfo {year} {1998})}\BibitemShut {NoStop}%
\bibitem [{\citenamefont {Vanier}(2005)}]{Vanier_2005}%
  \BibitemOpen
  \bibfield  {author} {\bibinfo {author} {\bibfnamefont {J.}~\bibnamefont {Vanier}},\ }\bibfield  {title} {\bibinfo {title} {Atomic clocks based on coherent population trapping: a review},\ }\href {https://doi.org/10.1007/s00340-005-1905-3} {\bibfield  {journal} {\bibinfo  {journal} {Applied Physics B}\ }\textbf {\bibinfo {volume} {81}},\ \bibinfo {pages} {421–442} (\bibinfo {year} {2005})}\BibitemShut {NoStop}%
\bibitem [{\citenamefont {Fleischhauer}\ \emph {et~al.}(2005)\citenamefont {Fleischhauer}, \citenamefont {Imamoglu},\ and\ \citenamefont {Marangos}}]{Fleischhauer_Imamoglu_Marangos_2005}%
  \BibitemOpen
  \bibfield  {author} {\bibinfo {author} {\bibfnamefont {M.}~\bibnamefont {Fleischhauer}}, \bibinfo {author} {\bibfnamefont {A.}~\bibnamefont {Imamoglu}},\ and\ \bibinfo {author} {\bibfnamefont {J.~P.}\ \bibnamefont {Marangos}},\ }\bibfield  {title} {\bibinfo {title} {Electromagnetically induced transparency: Optics in coherent media},\ }\href {https://doi.org/10.1103/RevModPhys.77.633} {\bibfield  {journal} {\bibinfo  {journal} {Reviews of Modern Physics}\ }\textbf {\bibinfo {volume} {77}},\ \bibinfo {pages} {633–673} (\bibinfo {year} {2005})}\BibitemShut {NoStop}%
\bibitem [{\citenamefont {Fourcault}\ \emph {et~al.}(2021)\citenamefont {Fourcault}, \citenamefont {Romain}, \citenamefont {Gal}, \citenamefont {Bertrand}, \citenamefont {Josselin}, \citenamefont {Prado}, \citenamefont {Labyt},\ and\ \citenamefont {Palacios-Laloy}}]{Fourcault_Romain_Gal_Bertrand_Josselin_Prado_Labyt_Palacios-Laloy_2021}%
  \BibitemOpen
  \bibfield  {author} {\bibinfo {author} {\bibfnamefont {W.}~\bibnamefont {Fourcault}}, \bibinfo {author} {\bibfnamefont {R.}~\bibnamefont {Romain}}, \bibinfo {author} {\bibfnamefont {G.~L.}\ \bibnamefont {Gal}}, \bibinfo {author} {\bibfnamefont {F.}~\bibnamefont {Bertrand}}, \bibinfo {author} {\bibfnamefont {V.}~\bibnamefont {Josselin}}, \bibinfo {author} {\bibfnamefont {M.~L.}\ \bibnamefont {Prado}}, \bibinfo {author} {\bibfnamefont {E.}~\bibnamefont {Labyt}},\ and\ \bibinfo {author} {\bibfnamefont {A.}~\bibnamefont {Palacios-Laloy}},\ }\bibfield  {title} {\bibinfo {title} {Helium-4 magnetometers for room-temperature biomedical imaging: toward collective operation and photon-noise limited sensitivity},\ }\href {https://doi.org/10.1364/OE.420031} {\bibfield  {journal} {\bibinfo  {journal} {Optics Express}\ }\textbf {\bibinfo {volume} {29}},\ \bibinfo {pages} {14467–14475} (\bibinfo {year} {2021})}\BibitemShut {NoStop}%
\bibitem [{\citenamefont {Meraki}\ \emph {et~al.}(2023)\citenamefont {Meraki}, \citenamefont {Elson}, \citenamefont {Ho}, \citenamefont {Akbar}, \citenamefont {Kozbial}, \citenamefont {Kolodynski},\ and\ \citenamefont {Jensen}}]{Meraki_Elson_Ho_Akbar_Kozbial_Kolodynski_Jensen_2023}%
  \BibitemOpen
  \bibfield  {author} {\bibinfo {author} {\bibfnamefont {A.}~\bibnamefont {Meraki}}, \bibinfo {author} {\bibfnamefont {L.}~\bibnamefont {Elson}}, \bibinfo {author} {\bibfnamefont {N.}~\bibnamefont {Ho}}, \bibinfo {author} {\bibfnamefont {A.}~\bibnamefont {Akbar}}, \bibinfo {author} {\bibfnamefont {M.}~\bibnamefont {Kozbial}}, \bibinfo {author} {\bibfnamefont {J.}~\bibnamefont {Kolodynski}},\ and\ \bibinfo {author} {\bibfnamefont {K.}~\bibnamefont {Jensen}},\ }\bibfield  {title} {\bibinfo {title} {Zero-field optical magnetometer based on spin alignment},\ }\href {https://doi.org/https://doi.org/10.1103/PhysRevA.108.062610} {\bibfield  {journal} {\bibinfo  {journal} {Physical Review A}\ }\textbf {\bibinfo {volume} {108}},\ \bibinfo {pages} {062610} (\bibinfo {year} {2023})}\BibitemShut {NoStop}%
\bibitem [{\citenamefont {Wang}\ \emph {et~al.}(2021)\citenamefont {Wang}, \citenamefont {Wu}, \citenamefont {Xiao}, \citenamefont {Wang}, \citenamefont {Peng},\ and\ \citenamefont {Guo}}]{Wang_Wu_Xiao_Wang_Peng_Guo_2021}%
  \BibitemOpen
  \bibfield  {author} {\bibinfo {author} {\bibfnamefont {H.}~\bibnamefont {Wang}}, \bibinfo {author} {\bibfnamefont {T.}~\bibnamefont {Wu}}, \bibinfo {author} {\bibfnamefont {W.}~\bibnamefont {Xiao}}, \bibinfo {author} {\bibfnamefont {H.}~\bibnamefont {Wang}}, \bibinfo {author} {\bibfnamefont {X.}~\bibnamefont {Peng}},\ and\ \bibinfo {author} {\bibfnamefont {H.}~\bibnamefont {Guo}},\ }\bibfield  {title} {\bibinfo {title} {Dual-mode dead-zone-free double-resonance alignment-based magnetometer},\ }\href {https://doi.org/10.1103/PhysRevApplied.15.024033} {\bibfield  {journal} {\bibinfo  {journal} {Physical Review Applied}\ }\textbf {\bibinfo {volume} {15}},\ \bibinfo {pages} {024033} (\bibinfo {year} {2021})}\BibitemShut {NoStop}%
\bibitem [{\citenamefont {Breschi}\ and\ \citenamefont {Weis}(2012)}]{Breschi_Weis_2012}%
  \BibitemOpen
  \bibfield  {author} {\bibinfo {author} {\bibfnamefont {E.}~\bibnamefont {Breschi}}\ and\ \bibinfo {author} {\bibfnamefont {A.}~\bibnamefont {Weis}},\ }\bibfield  {title} {\bibinfo {title} {Ground-state hanle effect based on atomic alignment},\ }\href {https://doi.org/10.1103/PhysRevA.86.053427} {\bibfield  {journal} {\bibinfo  {journal} {Physical Review A}\ }\textbf {\bibinfo {volume} {86}},\ \bibinfo {pages} {053427} (\bibinfo {year} {2012})}\BibitemShut {NoStop}%
\bibitem [{\citenamefont {Rochester}\ \emph {et~al.}(2012)\citenamefont {Rochester}, \citenamefont {Ledbetter}, \citenamefont {Zigdon}, \citenamefont {Wilson-Gordon},\ and\ \citenamefont {Budker}}]{Rochester_Ledbetter_Zigdon_Wilson-Gordon_Budker_2012}%
  \BibitemOpen
  \bibfield  {author} {\bibinfo {author} {\bibfnamefont {S.}~\bibnamefont {Rochester}}, \bibinfo {author} {\bibfnamefont {M.}~\bibnamefont {Ledbetter}}, \bibinfo {author} {\bibfnamefont {T.}~\bibnamefont {Zigdon}}, \bibinfo {author} {\bibfnamefont {A.}~\bibnamefont {Wilson-Gordon}},\ and\ \bibinfo {author} {\bibfnamefont {D.}~\bibnamefont {Budker}},\ }\bibfield  {title} {\bibinfo {title} {Orientation-to-alignment conversion and spin squeezing},\ }\href {https://doi.org/https://doi.org/10.1103/PhysRevA.85.022125} {\bibfield  {journal} {\bibinfo  {journal} {Physical Review A—Atomic, Molecular, and Optical Physics}\ }\textbf {\bibinfo {volume} {85}},\ \bibinfo {pages} {022125} (\bibinfo {year} {2012})}\BibitemShut {NoStop}%
\bibitem [{\citenamefont {LeGal}\ \emph {et~al.}(2019)\citenamefont {LeGal}, \citenamefont {Lieb}, \citenamefont {Beato}, \citenamefont {Jager}, \citenamefont {Gilles},\ and\ \citenamefont {Palacios-Laloy}}]{LeGal_Lieb_Beato_Jager_Gilles_Palacios-Laloy_2019}%
  \BibitemOpen
  \bibfield  {author} {\bibinfo {author} {\bibfnamefont {G.}~\bibnamefont {LeGal}}, \bibinfo {author} {\bibfnamefont {G.}~\bibnamefont {Lieb}}, \bibinfo {author} {\bibfnamefont {F.}~\bibnamefont {Beato}}, \bibinfo {author} {\bibfnamefont {T.}~\bibnamefont {Jager}}, \bibinfo {author} {\bibfnamefont {H.}~\bibnamefont {Gilles}},\ and\ \bibinfo {author} {\bibfnamefont {A.}~\bibnamefont {Palacios-Laloy}},\ }\bibfield  {title} {\bibinfo {title} {Dual-axis hanle magnetometer based on atomic alignment with a single optical access},\ }\href {https://doi.org/10.1103/PhysRevApplied.12.064010} {\bibfield  {journal} {\bibinfo  {journal} {Physical Review Applied}\ }\textbf {\bibinfo {volume} {12}},\ \bibinfo {pages} {064010} (\bibinfo {year} {2019})}\BibitemShut {NoStop}%
\bibitem [{\citenamefont {LeGal}\ and\ \citenamefont {Palacios-Laloy}(2022)}]{LeGal_Palacios-Laloy_2022}%
  \BibitemOpen
  \bibfield  {author} {\bibinfo {author} {\bibfnamefont {G.}~\bibnamefont {LeGal}}\ and\ \bibinfo {author} {\bibfnamefont {A.}~\bibnamefont {Palacios-Laloy}},\ }\bibfield  {title} {\bibinfo {title} {Zero-field magnetometry based on the combination of atomic orientation and alignment},\ }\href {https://doi.org/https://doi.org/10.1103/PhysRevA.105.043114} {\bibfield  {journal} {\bibinfo  {journal} {Physical Review A}\ }\textbf {\bibinfo {volume} {105}},\ \bibinfo {pages} {043114} (\bibinfo {year} {2022})}\BibitemShut {NoStop}%
\bibitem [{\citenamefont {Budker}\ \emph {et~al.}(2002)\citenamefont {Budker}, \citenamefont {Gawlik}, \citenamefont {Kimball}, \citenamefont {Rochester}, \citenamefont {Yashchuk},\ and\ \citenamefont {Weis}}]{Budker_Gawlik_Kimball_Rochester_Yashchuk_Weis_2002}%
  \BibitemOpen
  \bibfield  {author} {\bibinfo {author} {\bibfnamefont {D.}~\bibnamefont {Budker}}, \bibinfo {author} {\bibfnamefont {W.}~\bibnamefont {Gawlik}}, \bibinfo {author} {\bibfnamefont {D.~F.}\ \bibnamefont {Kimball}}, \bibinfo {author} {\bibfnamefont {S.~M.}\ \bibnamefont {Rochester}}, \bibinfo {author} {\bibfnamefont {V.~V.}\ \bibnamefont {Yashchuk}},\ and\ \bibinfo {author} {\bibfnamefont {A.}~\bibnamefont {Weis}},\ }\bibfield  {title} {\bibinfo {title} {Resonant nonlinear magneto-optical effects in atoms},\ }\href {https://doi.org/10.1103/RevModPhys.74.1153} {\bibfield  {journal} {\bibinfo  {journal} {Reviews of Modern Physics}\ }\textbf {\bibinfo {volume} {74}},\ \bibinfo {pages} {1153–1201} (\bibinfo {year} {2002})}\BibitemShut {NoStop}%
\bibitem [{\citenamefont {Blum}(2012)}]{Blum_2012}%
  \BibitemOpen
  \bibfield  {author} {\bibinfo {author} {\bibfnamefont {K.}~\bibnamefont {Blum}},\ }\bibinfo {title} {Irreducible components of the density matrix},\ in\ \href {https://doi.org/10.1007/978-3-642-20561-3_4} {\emph {\bibinfo {booktitle} {Density Matrix Theory and Applications}}},\ \bibinfo {editor} {edited by\ \bibinfo {editor} {\bibfnamefont {K.}~\bibnamefont {Blum}}}\ (\bibinfo  {publisher} {Springer},\ \bibinfo {address} {Berlin, Heidelberg},\ \bibinfo {year} {2012})\ p.\ \bibinfo {pages} {115–163}\BibitemShut {NoStop}%
\bibitem [{\citenamefont {Omont}(1977)}]{Omont_1977}%
  \BibitemOpen
  \bibfield  {author} {\bibinfo {author} {\bibfnamefont {A.}~\bibnamefont {Omont}},\ }\bibfield  {title} {\bibinfo {title} {Irreducible components of the density matrix. application to optical pumping},\ }\href {https://doi.org/10.1016/0079-6727(79)90003-X} {\bibfield  {journal} {\bibinfo  {journal} {Progress in Quantum Electronics}\ }\textbf {\bibinfo {volume} {5}},\ \bibinfo {pages} {69–138} (\bibinfo {year} {1977})}\BibitemShut {NoStop}%
\bibitem [{\citenamefont {Petrenko}\ \emph {et~al.}(2024)\citenamefont {Petrenko}, \citenamefont {Pazgalev},\ and\ \citenamefont {Vershovskii}}]{Petrenko_Pazgalev_Vershovskii_2024}%
  \BibitemOpen
  \bibfield  {author} {\bibinfo {author} {\bibfnamefont {M.~V.}\ \bibnamefont {Petrenko}}, \bibinfo {author} {\bibfnamefont {A.~S.}\ \bibnamefont {Pazgalev}},\ and\ \bibinfo {author} {\bibfnamefont {A.~K.}\ \bibnamefont {Vershovskii}},\ }\bibfield  {title} {\bibinfo {title} {A method of laser frequency stabilization based on the effect of linear dichroism in alkali metal vapors in a modulated transverse magnetic field},\ }\href {https://doi.org/10.3390/photonics11100926} {\bibfield  {journal} {\bibinfo  {journal} {Photonics}\ }\textbf {\bibinfo {volume} {11}},\ \bibinfo {pages} {926} (\bibinfo {year} {2024})}\BibitemShut {NoStop}%
\bibitem [{\citenamefont {Fomin}\ \emph {et~al.}(2025)\citenamefont {Fomin}, \citenamefont {Kozlov}, \citenamefont {Petrov}, \citenamefont {Smirnov}, \citenamefont {Petrenko},\ and\ \citenamefont {Zapasskii}}]{Fomin_Kozlov_Petrov_Smirnov_Petrenko_Zapasskii_2025}%
  \BibitemOpen
  \bibfield  {author} {\bibinfo {author} {\bibfnamefont {A.~A.}\ \bibnamefont {Fomin}}, \bibinfo {author} {\bibfnamefont {G.~G.}\ \bibnamefont {Kozlov}}, \bibinfo {author} {\bibfnamefont {M.~Y.}\ \bibnamefont {Petrov}}, \bibinfo {author} {\bibfnamefont {D.~S.}\ \bibnamefont {Smirnov}}, \bibinfo {author} {\bibfnamefont {M.~V.}\ \bibnamefont {Petrenko}},\ and\ \bibinfo {author} {\bibfnamefont {V.~S.}\ \bibnamefont {Zapasskii}},\ }\bibfield  {title} {\bibinfo {title} {Magnetic quadrupole dichroism in an isotropic medium},\ }\href {https://doi.org/10.1103/PhysRevA.111.023502} {\bibfield  {journal} {\bibinfo  {journal} {Phys. Rev. A}\ }\textbf {\bibinfo {volume} {111}},\ \bibinfo {pages} {023502} (\bibinfo {year} {2025})}\BibitemShut {NoStop}%
\bibitem [{\citenamefont {Happer}(1972)}]{Happer_1972}%
  \BibitemOpen
  \bibfield  {author} {\bibinfo {author} {\bibfnamefont {W.}~\bibnamefont {Happer}},\ }\bibfield  {title} {\bibinfo {title} {Optical pumping},\ }\href {https://doi.org/10.1103/RevModPhys.44.169} {\bibfield  {journal} {\bibinfo  {journal} {Reviews of Modern Physics}\ }\textbf {\bibinfo {volume} {44}},\ \bibinfo {pages} {169–249} (\bibinfo {year} {1972})}\BibitemShut {NoStop}%
\bibitem [{\citenamefont {Sobelman}(2012)}]{Sobelman_2012}%
  \BibitemOpen
  \bibfield  {author} {\bibinfo {author} {\bibfnamefont {I.~I.}\ \bibnamefont {Sobelman}},\ }  {\emph {\bibinfo {title} {Atomic spectra and radiative transitions}}},\ Vol.~\bibinfo {volume} {12}\ (\bibinfo  {publisher} {Springer Science \& Business Media},\ \bibinfo {year} {2012})\BibitemShut {NoStop}%
\bibitem [{\citenamefont {Varshalovich}\ \emph {et~al.}(1988)\citenamefont {Varshalovich}, \citenamefont {Moskalev},\ and\ \citenamefont {Khersonskii}}]{Varshalovich_Moskalev_Khersonskii_1988}%
  \BibitemOpen
  \bibfield  {author} {\bibinfo {author} {\bibfnamefont {D.}~\bibnamefont {Varshalovich}}, \bibinfo {author} {\bibfnamefont {A.}~\bibnamefont {Moskalev}},\ and\ \bibinfo {author} {\bibfnamefont {V.}~\bibnamefont {Khersonskii}},\ }\href {https://library.oapen.org/handle/20.500.12657/50493} {\emph {\bibinfo {title} {Quantum theory of angular momentum}}}\ (\bibinfo  {publisher} {World scientific},\ \bibinfo {year} {1988})\BibitemShut {NoStop}%
\bibitem [{\citenamefont {Budker}\ \emph {et~al.}(2004)\citenamefont {Budker}, \citenamefont {Kimball},\ and\ \citenamefont {DeMille}}]{Budker_Kimball_DeMille_2004}%
  \BibitemOpen
  \bibfield  {author} {\bibinfo {author} {\bibfnamefont {D.}~\bibnamefont {Budker}}, \bibinfo {author} {\bibfnamefont {D.~F.}\ \bibnamefont {Kimball}},\ and\ \bibinfo {author} {\bibfnamefont {D.~P.}\ \bibnamefont {DeMille}},\ }{\emph {\bibinfo {title} {Atomic physics: an exploration through problems and solutions}}}\ (\bibinfo  {publisher} {Oxford University Press, USA},\ \bibinfo {year} {2004})\BibitemShut {NoStop}%
\bibitem [{\citenamefont {Popov}\ \emph {et~al.}(2018)\citenamefont {Popov}, \citenamefont {Bobrikova}, \citenamefont {Voskoboinikov}, \citenamefont {Barantsev}, \citenamefont {Ustinov}, \citenamefont {Litvinov}, \citenamefont {Vershovskii}, \citenamefont {Dmitriev}, \citenamefont {Kartoshkin}, \citenamefont {Pazgalev},\ and\ \citenamefont {Petrenko}}]{Popov_Bobrikova_Voskoboinikov_Barantsev_Ustinov_Litvinov_Vershovskii_Dmitriev_Kartoshkin_Pazgalev_2018}%
  \BibitemOpen
  \bibfield  {author} {\bibinfo {author} {\bibfnamefont {E.~N.}\ \bibnamefont {Popov}}, \bibinfo {author} {\bibfnamefont {V.~A.}\ \bibnamefont {Bobrikova}}, \bibinfo {author} {\bibfnamefont {S.~P.}\ \bibnamefont {Voskoboinikov}}, \bibinfo {author} {\bibfnamefont {K.~A.}\ \bibnamefont {Barantsev}}, \bibinfo {author} {\bibfnamefont {S.~M.}\ \bibnamefont {Ustinov}}, \bibinfo {author} {\bibfnamefont {A.~N.}\ \bibnamefont {Litvinov}}, \bibinfo {author} {\bibfnamefont {A.~K.}\ \bibnamefont {Vershovskii}}, \bibinfo {author} {\bibfnamefont {S.~P.}\ \bibnamefont {Dmitriev}}, \bibinfo {author} {\bibfnamefont {V.~A.}\ \bibnamefont {Kartoshkin}}, \bibinfo {author} {\bibfnamefont {A.~S.}\ \bibnamefont {Pazgalev}},\ and\ \bibinfo {author} {\bibfnamefont {M.~V.}\ \bibnamefont {Petrenko}},\ }\bibfield  {title} {\bibinfo {title} {Features of the formation of the spin polarization of an alkali metal at the resolution of hyperfine sublevels in the 2s1/2 state},\ }\href {https://doi.org/10.1134/S0021364018200122} {\bibfield
  {journal} {\bibinfo  {journal} {JETP Letters}\ }\textbf {\bibinfo {volume} {108}},\ \bibinfo {pages} {513–518} (\bibinfo {year} {2018})}\BibitemShut {NoStop}%
\bibitem [{\citenamefont {Alexandrov}\ \emph {et~al.}(1993)\citenamefont {Alexandrov}, \citenamefont {Chaika},\ and\ \citenamefont {Khvostenko}}]{Alexandrov_Chaika_Khvostenko_1993}%
  \BibitemOpen
  \bibfield  {author} {\bibinfo {author} {\bibfnamefont {E.~B.}\ \bibnamefont {Alexandrov}}, \bibinfo {author} {\bibfnamefont {M.~P.}\ \bibnamefont {Chaika}},\ and\ \bibinfo {author} {\bibfnamefont {G.~I.}\ \bibnamefont {Khvostenko}},\ }  {\emph {\bibinfo {title} {Interference of atomic states}}},\ Vol.~\bibinfo {volume} {7}\ (\bibinfo  {publisher} {Springer},\ \bibinfo {year} {1993})\BibitemShut {NoStop}%
\end{thebibliography}

\end{document}